\documentclass[english]{article}
\usepackage[T1]{fontenc}
\usepackage[latin9]{inputenc}
\usepackage{geometry}
\usepackage{amsmath}
\usepackage{graphicx}



\usepackage{babel}
\usepackage{csquotes}
\title{A Viscoelastic Model for Droplet Breakup in Dense Emulsions}
\author{ Joseph D. Peterson$^1$, Ioannis Bagkeris$^2$, Vipin Michael$^3$}
\date{$^1$Henry Samueli School of Engineering, UCLA, Boelter Hall, 4531J \\
         $^2$Unilever R\&D, Port Sunlight Laboratory, Quarry Road East, Bebington, Wirral CH63 3JW, UK \\
         $^3$School of Engineering, University of Manchester, Manchester, M13 9PL, UK \\
         \today
         }
\begin{document}
\maketitle
\begin{abstract}
When processing dense emulsions, complex flows stretch and deform
droplets to the point of breakup, changing the droplet size distribution
and the mechanical properties of the final product. For steady homogeneous
flows, a droplet's shape and proclivity to breakup can be inferred
from a Capillary number, comparing the strain rate with the typical
shape relaxation time. More generally, a droplet's shape depends on
its whole history of deformation, and this distinction can be salient
under conditions relevant to industrial processing. Here, we develop
a coupled population balance and droplet shape evolution model ``STEPB''
with a breaking rate determined by the droplet's shape. In a first
application of the STEPB model, we consider droplet breakup during
relaxation of step shear, where we predict non-trivial strain-dependent
final daughter droplet distributions. We also compare predictions
from STEPB against experimental observations in steady flow conditions,
illuminating key weaknesses in the model due to neglected terms.

\global\long\def\ppt{\frac{\partial}{\partial t}}%

\global\long\def\Tb{\boldsymbol{\Theta}}%

\global\long\def\ub{\boldsymbol{u}}%

\global\long\def\kz{\boldsymbol{\kappa}_{\zeta}}%
\end{abstract}

\section{Introduction}

Dense emulsions are stabilized mixtures of immiscible or partially
miscible fluids, comprising a dispersed droplet phase and a continuous
fluid phase. Stabilization of an emulsion is commonly attained by
the introduction of interfacially active components, such as surfactants,
proteins, or solid particles, that create a steric or thermodynamic
barrier to droplet coalescence \cite{tadros2016emulsions,pickering1907cxcvi,mcclements2004protein}.
As a platform for consumer product formulation, dense emulsions are
found in applications ranging from mayonnaise to cosmetic/pharmaceutical
creams.

In principle, dense emulsions are not difficult to produce: for example,
a simple mayonnaise is made by combining oil, vinegar, and egg yolks
in appropriate proportions and blending vigorously \cite{harrison1985factors}.
In practice, however, a profitable implementation at industrial scales
must be precise, robust, and economical - when addressing these constraints,
the true complexities of emulsion processing become apparent. For
example, what is the most energy-efficient way to ``vigorously blend''
two immiscible fluids? Can one devise a combined formulation and processing
strategy that reduces the fraction of costly ingredients (oil, egg)
while preserving the same stability, taste, and ``spreadability''
in the final product? In pursuit of such questions, iterations on
pilot-scale processing equipment are expensive and time-consuming;
therefore, computational and modeling tools have an important role
to play, reducing the design-space of interest and accelerating the
overall timeline for innovation and deployment of new ideas.

In this work, we will focus on modeling tools that relate processing
history to the final droplet size distribution, as might be needed
to study ``vigorous blending'' steps. When formulating dense emulsions,
one typically begins from a coarse pre-mix with a population of large
droplets, after which complex flows stretch and deform the droplets
to induce breakup and establish a final population of smaller droplets
\cite{dubbelboer2016pilot,maindarkar2014prediction}. The full details
of such a process can be overwhelmingly complex, but here we will
outline three existing modeling strategies that provide insights to
the relationship between an imposed deformation and the resultant
changes in the droplet shape and proclivity to breakup.

Some of the oldest work on the nonlinear rheology of dilute emulsions
dates to Taylor \cite{taylor1934formation}, who developed a pertubation-expansion
approach to model changes in droplet shape at low strain rates, $Ca\ll1$,
where the Capillary number $Ca$ compares typical viscous stresses
and interfacial stresses. Research following this approach has since
accounted for higher-order corrections \cite{barthes1973deformation,rallison1984deformation}
and a broader library of interfacial physics \cite{narsimhan2019shape, singh2020deformation}.
This vein of research explicitly tracks an approximate representation
of the droplet shape evolution, as encoded by a set of symmetric tensors,
and the results are indeed useful for understanding the behavior of
dilute emulsions at low Capillary numbers. In principle, low-$Ca$
expansions are not suited to describe droplet breakup, but when such
theories are extrapolated to $Ca\sim O(1)$ one can find indirect
evidence of droplet breakup in terms of a ``turning point'' in the
set of steady flow solutions \cite{barthes1973deformation, singh2020deformation}.
In considering dense emulsions, however, the rigor of a perturbation-expansion
approach is not feasible, mostly because dense emulsions do not have
well-defined flow-fields at the droplet-scale; in heterogeneous materials,
one generally expects that individual particles/droplets can experience
different local strain histories for the same bulk deformation \cite{degiuli2017unifying}.
Thus, whereas droplet breakup is a deterministic function of strain
history in dilute emulsions, it can be viewed as partly stochastic
in dense emulsions.

To avoid the complications of explicitly tracking the size and shape
of individual droplets, Doi and Ohta (DO) developed a remarkably simple
viscoelastic model that achieved closure with only the total droplet
stress and the total droplet surface area \cite{doi1991dynamics}.
The DO model has no conceptual difficulty with dense emulsions or
high $Ca$, and the complex details of droplet breakup and coalescence
are simply handled implicitly through the relaxation of stress and
surface area. Unfortunately, however, this model has conceptual difficulties
with stabilized emulsions, since it is a scale-free model suitable
only for systems with neat interfaces where coalescence is not arrested.
Additionally, because the DO model avoids tracking the droplet size
distribution, it may be less useful in applications where processing
objectives are defined in terms of a target droplet size distribution.

Broadly speaking, the above approaches are carefully-derived from
well-defined microscopic models, and this rigor naturally gives rise
to a more restricted range of application. For a more flexible set
of tools with less restricted descriptive power, one can consider
more phenomenological models. Neglecting the viscoelasticity of the
droplets, for example, one can often infer the shape of a droplet
(and its typical time to break) from the instantaneous strain rate
\cite{wieringa1996droplet}. Likewise, the stochasticity of droplet
breakup in dense emulsions can be approximated via a Poisson process
in a simple first-order rate expression. Dramatic simplifications
of this nature still prove insightful for many applications, and the
resulting models are simple enough and flexible enough for practical
use in complex flow calculations \cite{dubbelboer2016pilot,maindarkar2014prediction}.
A closely related simplification strategy beyond the scope of our
present interests can also be found in breakage kernels applied in
Reynolds-averaged models of turbulent flow. There, the breaking rate
is described in terms of the mean turbulent energy dissipation rate
\cite{coulaloglou1977description}. This is an appropriate simplification
scheme for Reynolds-averaged models, but not helpful for laminar flows
or fully-resolved turbulent flows.

In our view, the principal weakness of existing phenomenological models
lies in the loss of viscoelasticity and, more precisely, the loss
of an explicit evolution equation pertaining to the droplet shape/stress.
The shape of a droplet - as well as its stress and proclivity to breakup
- depends not on an instantaneous strain rate but on its entire history
of strain. This principle has been clearly demonstrated in the dilute
emulsions literature, where Stone et. al. studied the breakup of droplets
\emph{after} cessation of flow \cite{stone1986experimental}. To our
knowledge, the history-dependence of droplet breakup has yet to be
demonstrated directly in experiments on dense emulsions, but here
we proceed on the assumption that a connection between droplet breakup
and droplet deformation history will exist irrespective of whether
the surrounding media is heterogeneous or homogeneous. That being
said, it is our view that the details of droplet breakup in dense
and dilute emulsions are substantial and will require different mathematical
machinery built on different assumptions and abstractions.

Given the abundance of detailed fundamental work on dilute emulsions
and the relative dearth of data on droplet breakup dynamics in dense
emulsions, it might seem that dilute emulsions present the more natural
starting point for model development. However, we argue that for the
specific application at hand - coupling rheology and droplet size
evolution - it is more natural to begin with dense emulsions. There
are many reasons for this, but they all seem to stem from the fact
that deterministic breakup events in dilute emulsions present a significant
complication for population balance equations, where Poisson-distributed
processes are more naturally accommodated. But is it any more reasonable
to treat the ``stochastic'' breakup processes in dense emulsions
as Poisson distributed? We think so, but this is really an open question
that (to our knowledge) has not even been asked until now. Indeed,
absent a useful interpretive modeling framework there has been very
little incentive for experiments that study the statistics and dynamics
of droplet breakup in dense emulsions. There are several such approximations
and abstractions presented in this report, and it is our hope that
these assumptions will not go unchallenged but will instead motivate
a focused experimental inquiry.

The outline of our modeling approach is summarized as follows: in
keeping with the phenomenological approach, favoring flexibility and
simplicity over rigor, we will rely on simple ellipsoidal droplet
shapes to interpolate properties of the complex non-ellipsoidal structures
encountered in droplet breakup. For a collection of droplets with
the same volume, we will compute an average ellipsoidal representation
of the droplet shape within that population. When the average droplet
shape is sufficiently deformed, breakup proceeds as a Poisson process.
The daughter droplets are assumed to be ellipsoidal with a shape consistent
with the stress held by the parent droplet. This picture is then extended
to all droplet sizes within a given distribution, yielding a set of
coupled population balance equations and shape evolution equations,
which we call the ``shape tensor emulsion population balance'' (STEPB)
model. Overall, we believe that the STEPB model yields a simple and
effective interpolation of the deformation-history-dependent physics
relevant to droplet breakup in dense emulsions. Existing models are
categorially unable to take on such a challenge, so while there are
many aspects of STEPB that remain improvable (e.g. including droplet
coalesence and/or sufractant mass transport effects), the progress
that we present here should be considered a major step forward.

Briefly, we will summarize two novel ideas that emerge from the STEPB
model. First, we will show that conserving stress during droplet breakup
creates complex strain-dependent daughter droplet distributions even
when individual breakup events are constrained to simple binary splitting.
This is possible because the daughter droplets themselves can be sufficiently
strained to break again (and again) when the initial imposed strain
is large. In our view, this is an important innovation because it
is impractical to accomodate history-dependent nuances of the daughter
droplet distributions into existing phenomenological models of dense
emulsions. Second, in defining an ``average'' shape for a collection
of equally-sized droplets, we provide the first physically-motivated
argument for the widely-used log-transformation of a constitutive
equation. The mathematical foundations for this transform were developed
for computational fluid dynamics (CFD) studies of polymeric materials,
as a way of avoiding numerical instabilities \cite{fattal2004constitutive}.
Here, however, we argue that the log-transformation is ideal for handling
conserved volume constraints in shape relaxation and shape averaging
operations.

The outline of this paper is as follows: In section \ref{sec:Governing-Equations},
we outline the bare framework of the STEPB model in terms of coupled
equations for momentum balance, droplet size distribution, and droplet
shape evolution. In section \ref{sec:Details-of-the}, we continue
with more implementation-specific details, relating the ellipsoidal
droplet shape tensor to an estimated droplet surface area, stress,
and breaking rate. In section \ref{sec:step-shear}, we discuss model
predictions for monodisperse emulsions in step-shear and outline a
strategy for parameterizing the model through a confrontation with
experimental data. In section \ref{sec:experiments}, we compare against
experimental data for polydisperse emulsions in flows approximating
steady simple shear, demonstrating areas of weakness in the current
implementation. Finally, section \ref{sec:Summary-and-Conclusions}
closes the report by summarizing key ideas, results, and directions
for future research.

\section{\label{sec:Governing-Equations}Governing Equations}

The principle objective of this report is to introduce simplified
equations that broadly capture important aspects of the relationship
between droplet size distribution and processing history, especially
as it pertains to unsteady, complex flow conditions in which the shape
of a droplet cannot be reliably inferred from an instantaneous strain
rate \cite{stone1986experimental}.

To accomplish this, we begin with section \ref{subsec:Momentum-Balance-Equations}
where we present momentum balance equations, including a discussion
on elastic vs capillary stresses and the broader rheological context
for which our model has been designed. Then, in section \ref{subsec:Population-Balance-Equations}
we review the general framework of population balance equations (PBE)
for describing changes in a droplet size distribution. Finally, we
present a viscoelastic constitutive equation for the average droplet
shape as a function of droplet size, incorporating the effects of
deformation, relaxation, and droplet breakup. Finally, we discuss
how the viscous and capillary stresses enter into a momentum balance
equation.

This section provides a bare outline of the modeling framework itself,
and application-specific details (e.g. the coupling between droplet
shape and droplet stress, droplet breakup, etc.) are covered in section
\ref{sec:Details-of-the}. In separating the framework from its implementation,
we aim to make the model more open to revision and fine-tuning as
needed for future research and/or application-specific contexts. The
present implementation as outlined here will be referred to as the
``shape tensor emulsion population balance'' model, or STEPB. Additional
details relevant to numerical solutions of the STEPB model are covered
in appendix \ref{sec:Addition-Details-on}.

\subsection{\label{subsec:Momentum-Balance-Equations}Momentum Balance Equations}

The rheology of dense suspensions can be classified into at least
three distinctive regimes, with the most well-understood regimes being
perhaps the least relevant to this work. Thus it is worth taking a
moment to identify the relevant rheological context for our present
study.

At very low shear rates, the first distinction to consider is whether
or not droplets are packed densely enough to create a yield stress
material \cite{scheffold2013linear}. When the yield stress is below
the capillary modulus of a typical droplet (as is especially true
for systems just above jamming), droplets remain largely un-deformed
and are not susceptible to breakup along the yield surface.

As the shear rate increases, both shear stresses and normal stresses
increase in a power-law relation \cite{khabaz2020particle}, with
similar power law scalings for both shear stresses and normal stresses.
Here again, viscous stresses are well below the capillary modulus
of a typical droplet, and droplet breakup should not be expected.

At higher shear rates, there is evidence that the power law regime
transitions to a Newtonian regime \cite{caggioni2020variations},
but there is comparatively little data for direct observation of said
regime. The existence of at least one terminal Newtonian regime is
perhaps to be expected, but it is not obvious (1) what factors determine
the associated viscosity and (2) whether the indirectly-observed Newtonian
regime is indeed the terminal one.

In our work, we consider the third regime, beginning with a Newtonian
response, to be the only one relevant to industrial droplet milling
processes. The rheological response will be decomposed into two separate
contributions: an elastic contribution from droplet capillary stresses,
and a viscous response from the constituent Newtonian fluids themselves.
For simplicity, we will assume that the effective viscosity for the
purely viscous response is both (1) independent of the individual
droplet configurations and (2) representative of a ``background''
viscosity felt by droplets attempting to relax their shape. The estimation
and interpretation of such a background viscosity, which we call $\mu^{eff}$,
is beyond the scope of our present work but has been considered extensively
elsewhere \cite{faroughi2015generalized,pal2003viscous}, and any
of these existing ideas could be appended to our modeling framework
in application. We also note that $\mu^{eff}$is distinct from the
indirectly-measured Newtonian viscosity measured by Caggioni et. al.
\cite{caggioni2020variations}, since in principle the latter also
includes a contribution from the capillary stress response.

Turning now to the momentum balance equations, deformed droplets will
exert capillary stresses on the fluid. The total capillary stress
tensor $\boldsymbol{\boldsymbol{\sigma}_{T}}$ enters into the momentum
balance equation as:

\begin{equation}
\rho\Big[\ppt\ub+\ub\cdot\nabla\ub\Big]=-\nabla P+\rho\boldsymbol{g}+\nabla\cdot\boldsymbol{\boldsymbol{\sigma}_{T}}+\mu^{eff}\nabla^{2}\ub
\end{equation}

\begin{equation}
\nabla\cdot\ub=0
\end{equation}

where $\ub$ is the bulk velocity, $\rho$ is the fluid density, $P$
is the pressure, $\boldsymbol{g}$ is the gravitational field, $\mu^{eff}$
is the effective viscosity of the fluid accounting for the total viscous
dissipation in both the dispersed and continuous phases. For dense
suspensions, we assume that capillary stresses are greater than gravitational
forces, such that the suspension exhibits a uniform fluid density
and the prospects of droplet settling or gel collapse can be ignored.

The capillary stresses must ensure that the reversible work needed
to deform a droplet is equal the change in free energy induced by
said deformation - this thermodynamic constraint will be used to specify
the relationship between a shape tensor $\Tb$ and the stress tensor
$\boldsymbol{\sigma}$ once we have specified a functional form for
the droplet free energy $F(\Tb)$ in the next section of this work.

\subsection{\label{subsec:Population-Balance-Equations}Population Balance Equations}

Given a collection of droplets, we define a number density distribution
$n(\boldsymbol{x},t,v)\delta v$ as the number density of droplets
within a small control volume at position $\boldsymbol{x}$ and time
$t$ with droplet volume $v$ in the narrow interval of $v\in[v,v+\delta v]$.
For concise notation, the spatial and temporal dependencies of all
variables will be notationally suppressed in the equations that follow,
and only dependencies on droplet volume $v$ will be shown explicitly.

In the most general expression, the droplet size distribution changes
over time due to not only droplet breakup but also processes like
advection, diffusion, nucleation, growth, and droplet coalescence
\cite{ramkrishna2000population}:

\begin{equation}
\ppt n(v)+\nabla\cdot(\ub(v)n(v))=D(v)\nabla^{2}n(v)+\dot{n}_{B}(v)+\dot{n}_{C}(v)+\dot{n}_{N}(v)+\dot{n}_{G}(v)+\cdots\label{eq:general_PBE}
\end{equation}

In equation \ref{eq:general_PBE}, the second term on the left-hand
side (LHS) represents advection with a bulk velocity field $\ub(v)$
that may vary with particle size, and the first term on the right-hand
side (RHS) represents diffusion with a size-dependent diffusion coefficient
$D(v)$. The terms that follow, of which there may be more than those
explicitly ennumerated here, represent the effect of changes to the
droplets themselves: processes like breakup $\dot{n}_{B}$, coalescence
$\dot{n}_{C}$, nucleation $\dot{n}_{N}$, and growth $\dot{n}_{G}$.
In the limited focus of the present work, we are most interested in
understanding droplet breakup in unsteady flows - therefore, we will
only explicitly consider the breakup term $\dot{n}_{B}$. This restricted
focus is justifiable when thinking about surfactant-stabilized emulsions
with a significant excess of surfactant (e.g. mayonnaise). Likewise,
because we are not yet looking at complex flows the advection and
diffusion terms will also be suppressed.

Expanding on the breakup term, $\dot{n}_{B}(v)$ contains two separate
terms representing the outflux/influx of droplets with volume $v$
due to (1) the destruction of droplets with size $v$ via breakup
and (2) the production of droplets with volume v via the breakup of
larger droplets with volume $v'>v$:

\begin{equation}
\dot{n}_{B}=\dot{n}_{B,\text{in}}+\dot{n}_{B,\text{out}}
\end{equation}

The outflux of droplets due to breaking $\dot{n}_{B,\text{out}}$
must be proportional to the total number of droplets in the system,
$\dot{n}_{B}(v)\sim n(v)$ with a prefactor $g(v)$ that describes
a size-dependent breaking rate:

\begin{equation}
\dot{n}_{B,\text{out}}=-g(v)n(v)\label{eq:ndot_B_out}
\end{equation}

For a collection of equally-sized droplets, the rate at which droplets
in the group are breaking will depend on the present shape of all
individual droplets. Lacking such detailed information, however, one
can assume that the overall rate of breaking correlates to an ``averaged''
measure of the droplet shape - proposed details for such a correlation
will be fleshed out in section \ref{sec:Details-of-the}.

The influx term $\dot{n}_{B,\text{in}}$ integrates over all breaking
events occuring in droplets of size $v'>v$, weighted by the liklihood
$\beta(v,v')$ that a droplet of size $v$ would be produced following
breakup of a droplet with size $v'$:

\begin{equation}
\dot{n}_{B,\text{in}}=\int_{v}^{\infty}dv'\beta(v,v')g(v')n(v')\label{eq:ndot_B_in}
\end{equation}

The so-called daughter droplet distribution $\beta(v,v')$ can be
a major source of complexity in many PBE models of emulsions, since
the distribution changes depending on how the droplets were deformed
prior to breakup \cite{stone1986experimental}. In our approach, however,
we will explicitly track a measure of the droplet shape evolution,
and we will allow information about the state of droplet deformation
to be inherited by the daughter droplets themselves. Where the first
generation of daughter droplets is sufficiently deformed, a second
generation of daughter droplets may follow, even in the absence of
further deformation. This cascade of breakage eventually terminates,
as successive generations of daughter droplets will be less deformed.
In our view, the breakage cascade can be used to emulate some aspects
of deformation-dependent daughter droplet distributions, though we
acknowledge limitations with respect to capturing fine details like
satellite droplets.

In all, it is our view that ensuring stress conservation through the
breakage process partially alleviates the pressure to construct detailed
forms for $\beta(v,v')$, with the details of the breakage kernel
now being of secondary importance. Therefore, as a first implementation
we will suppose that breakup is always binary and symmetric, $\beta(v,v')=\delta(v-v'/2)$,
with every breakup producing just one pair of identical droplets.
Many readers may be unconvinced that our assumption of binary breakage
is as benign as we claim, given the importance of a carefully-constructed
$\beta(v,v')$ in other models for emulsion breakup. For these readers,
we have included calculations for a gerneralization built on a ternary
breakup assumption $\beta(v,v')=\delta(v-v'/3)$ in section \ref{subsec:Strain-Sweep-to},
and the differences are found to be small even when all other model
parameters are held fixed. Once again, binary and ternary breakage
approximations are both too crude to capture fine details like satellite
droplets, for example.

With a simplification to binary breakup, the integral in equation
\ref{eq:ndot_B_in} collapses and the overall PBE (neglecting all
processes except breakup) simplifies to:

\begin{equation}
\ppt n(v)=-g(v)n(v)+2g(2v)n(2v)\label{eq:ndot_final}
\end{equation}

Details of the numerical strategies we have used when solving equation
\ref{eq:ndot_final} can be found in appendix \ref{sec:Addition-Details-on}.
To summarize the appendix contents, we have employed two different
schemes depending on the choice of initial condition. Where the droplets
are initially monodisperse, we compute the population at discrete
droplet sizes corresponding to successive divisions of the original
droplet size. Where the droplets are initially polydisperse, we make
use of our recently developed ``method of the inverse cumulative
distribution function'', which is a flexible low-mode discretization
strategy that we believe is uniquely suited for solving population
balance equations in complex fluids. For further discussion, see appendix
\ref{sec:Addition-Details-on}.

\subsection{Droplet Configuration Equations}

In this section, we will develop equations that describe the evolution
of droplet shape in flow. First, section \ref{subsec:Droplet-Shape-and}
deals with defining a droplet shape and describing changes to that
shape induced by flow, momentarily ignoring capillary forces and population
balance dynamics. Next, section \ref{subsec:Constraints-on-shape}
describes the constraints on shape relaxation dynamics (an explicit
functional form will be chosen in section \ref{sec:Details-of-the}).
Finally, section \ref{subsec:shape_averaging} considers the average
shape evolution for a collection of droplets and the influence of
population balance dynamics.

\subsubsection{\label{subsec:Droplet-Shape-and}Droplet Shape and Deformation Response}

To describe how droplets change shape in response to deformation,
we need (1) an expression to define the droplet surface, and (2) an
equation to describe how that surface responds to flow. In the most
general form, the we can define the droplet surface as a manifold
in three-space satisfying $f(\boldsymbol{r})=0$. For motion of the
droplet surface, the rigorous approach involves solving momentum balance
equations inside/outside of the droplet, coupled by a kinematic boundary
condition. Here, we take a more phenomenological approach and assume
that droplets deform affinely with a bulk velocity gradient $\boldsymbol{\kappa}=\nabla\boldsymbol{u}$,
less some amount of slip:

\begin{equation}
\ppt\boldsymbol{r}=(\boldsymbol{\kappa}-\zeta\boldsymbol{E})\cdot\boldsymbol{r}\label{eq:surface_motion}
\end{equation}

where $\boldsymbol{E=}(\boldsymbol{\kappa}+\boldsymbol{\kappa}^{T})/2$
is the symmetric portion of the strain field and $\zeta\in[0,1]$
is the slip parameter. Finite slip may be important when the dispersed
phase is significantly more viscous than the continuous phase, for
example, increasing the critical shear rate for breakup (c.f. appendix
\ref{sec:Slip-and-the}). For concise notation, we define the effective
velocity gradient $\kz=\boldsymbol{\kappa}-\zeta\boldsymbol{E}$.

Next, we will consider a convenient functional form for the surface
manifold equation $f(\boldsymbol{r})=0$. Assuming the droplet has
an axis of 180 degree rotational symmetry about its center we can
define the droplet shape through a truncated expansion of tensors
with increasing rank:

\begin{equation}
f(\boldsymbol{r})=-1+\boldsymbol{A}:\boldsymbol{rr}+\boldsymbol{B}::\boldsymbol{rrrr}+\cdots\label{eq:surface_def}
\end{equation}

The tensors in this series are all symmetric to any exchange of indices,
so a rank-4 tensor needs just 15 unique entries. There are two advantages
of describing the droplet surface in this way, namely (1) the opportunity
to continuously resolve breakup into isolated droplets with higher-order
tensors and (2) the ease of producing evolution equations for shape
tensors when using the non-affine deformation rule of equation \ref{eq:surface_motion}.
With regard to the former consideration, note that there can be multiple
roots for the intersection of the droplet surface with an orientation
vector $\boldsymbol{n}=\boldsymbol{r}/|\boldsymbol{r}|$, which is
not possible if the surface is explicitly defined as a function of
orientation, $r(\boldsymbol{n}).$

Here, however, we will restrict ourselves to the assumption of an
ellipsoidally shaped droplet, truncating the tensor series at its
first term $\boldsymbol{A}$. This is a widely used approximation
for modeling non-breaking droplets in dilute emulsions \cite{minale2010models},
but some further discussion is required to interpret its use in dense
emulsions and emulsions where droplet breakup occurs.

Regarding dense emulsions, droplet shapes are generally never spherical
or ellipsoidal but instead faceted where droplets press together.
Here, we suggest that an ellipsoidal approximation is less intended
as a quantitative description of an individual droplet's present shape
and should instead be taken as a qualitative description of typical
deformations across a collection of droplets.

Regarding droplet breakup, the true process involves a progression
of droplet shapes with intermediate structures that cannot be understood
as ellipsoidal in any sense. All the same, we argue that a limited
set of relevant measures (e.g. surface area and stress) can be can
be effectively interpolated via a superposition of broken (ellipsoidal)
and unbroken (ellipsoidal) droplets. We are therefore leveraging droplet
breakup to capture both discrete topological transitions and continuous
non-ellipsoidal morphological changes. Over-leveraging droplet breakup
in this way is likely not appropriate for monodisperse dilute emulsions,
where droplet breakup always occurs at discrete moments in time and
highly non-ellipsoidal morphologies are possible even without droplet
breakup \cite{stone1986experimental}. For dense emulsions, however,
we expect that a broader distribution of droplet deformation histories
and breakup times makes it impractical or even unnecessary to clearly
separate the morphological and topological aspects of the breakup
process. For a cartoon schematic of how the ellipsoidal droplet approximation
is interpreted for highly distorted droplet shapes very near to breakup,
we refer the reader to Figure \ref{fig:droplet_cartoon} in section
\ref{subsec:Daughter-Droplet-Shapes} and the surrounding discussion.

In this paper, we can not yet provide a complete quantitative assessment
for the usefulness of our ellipsoidal approximation in the specific
context of dense emulsion droplet breakup - direct observation of
individual droplet shapes in dense emulsions under strong flow conditions
is a difficult challenge from both experimental and computational
approaches, and must be deferred to future work. Without the ability
to test this crucial approximation, the goal of this paper is to highlight
the promise and potential of the modeling framework that naturally
follows.

Continuing with our ellipsoidal droplet approximation, the tensor
$\boldsymbol{A}$ is symmetric, positive definite, rank-2 tensor whose
principle eigenvectors identify the axis of an ellipsoid, and whose
eigenvalues give the reciprocal square of the corresponding ellipsoid
dimensions. The evolution equation for $\boldsymbol{A}$ can be found
by applying a time derivative to equation \ref{eq:surface_def} along
the surface $f(r)=0$:

\begin{equation}
\ppt\boldsymbol{A=-u\cdot\nabla A-}\kz^{T}\cdot\boldsymbol{A}-\boldsymbol{A}\cdot\kz
\end{equation}

It is perhaps more useful to frame our shape tensor in terms of its
inverse, $\boldsymbol{C}=\boldsymbol{A}^{-1}/r_{0}(v)^{2}$, so that
the principle axis of our shape tensor correspond to the square (as
opposed to inverse square) of the axis of the ellipsoid it represents,
normalized by the equilibrium (spherical) droplet radius $r_{0}(v)$.
In this case, the evolution equation for $\boldsymbol{C}$ can be
written as:

\begin{equation}
\ppt\boldsymbol{C}=-\boldsymbol{u}\cdot\nabla\boldsymbol{C}+\boldsymbol{C}\cdot\kz^{T}+\kz\cdot\boldsymbol{C}\label{eq:GSD_def}
\end{equation}

This is the so-called Gordon-Schowalter derivative commonly used in
viscoelastic models of polymeric fluids \cite{larson2013constitutive}.
As a closing note to this section, the slip parameter $\zeta$ can
in principle be a function of the shape tensor itself - in studies
comparing to experimental results, this could be a useful means of
fine-tuning the simplistic picture given here.

\subsubsection{\label{subsec:Constraints-on-shape}Constraints on shape relaxation
dynamics}

Given an ellipsoidal droplet defined by a shape tensor $\boldsymbol{C}$
with equilibrium radius $r_{0}$, the volume of the droplet is given
by:

\begin{equation}
V=\frac{4}{3}\pi r_{0}^{3}\sqrt{\text{det}(\boldsymbol{C})}=\frac{4}{3}\pi r_{0}^{3}
\end{equation}

Note that $\det(\boldsymbol{C})\equiv1$ where the droplet is assumed
to be incompressible.

In the absence of flow, we will suppose that droplets relax back to
a spherical configuration, $\boldsymbol{C}=\boldsymbol{I}$, where
$\boldsymbol{I}$ is the identity tensor. This assumption would fail
for dense emulsions with a yield stress, but the yield surface is
generally well out-of-frame when emulsions are driven towards droplet
breakup (c.f. section \ref{subsec:Momentum-Balance-Equations}). It
is possible to amend our model to hold a finite yield stress by assuming
a configuration dependent relaxation time \cite{saramito2007new},
but this is outside the scope of our present interests.

As the droplets relax their shape towards equilibrium, we will first
assume that the rate of relaxation for a given droplet depends only
on the shape of the droplet in question:

\begin{equation}
\ppt\boldsymbol{C}=\cdots+\boldsymbol{R}(\boldsymbol{C})\label{eq:adding_relaxation}
\end{equation}

Here and elsewhere leading ellipses on the RHS denote terms previously
introduced, in this case advection and deformation terms from equation
\ref{eq:GSD_def}.

In seeking a functional form for the relaxation tensor $\boldsymbol{R}$,
there are two constraints to consider. First, as droplets relax their
shape their volume must be conserved at all times, $\det(\boldsymbol{C})=1,$
$\partial_{t}\text{det}(\boldsymbol{C})=0$, yielding the constraint:

\begin{equation}
\boldsymbol{C}^{-1}:\boldsymbol{R}(\boldsymbol{C})=0\label{eq:R_const_begin}
\end{equation}

Second, given a shape-dependent droplet free energy $F(\boldsymbol{C})$,
shape relaxation must decrease the droplet's free energy. Assuming
a constant surface tension, this simply becomes:

\begin{equation}
\frac{\delta F}{\delta\boldsymbol{C}}:R(\boldsymbol{C})\leq0
\end{equation}

Finally, shape relaxation must take place on a time-scale consistent
with the balance of viscous and capillary forces in the system. Let
us assume that viscous resistance to shape relaxation primarily comes
from the medium surrounding the droplet, with viscosity $\mu^{eff}$.
Capillary stresses will scale in proportion to the surface tension
$\Gamma$ and inversely with the droplet radius $r_{0}$. If the droplet
relaxes its shape on a timescale $\tau$, then viscous stresses will
be on the order of $\mu^{eff}/\tau$. Balancing viscous and capillary
stresses for a single droplet, we find that $\tau$ goes as:

\begin{equation}
\frac{1}{\tau}\sim\frac{\Gamma}{\mu^{eff}r_{0}}
\end{equation}

and so, considering only the timescale of stress relaxation (and not
its dependence on configuration) we find:

\begin{equation}
\boldsymbol{R}(\boldsymbol{C})\sim\frac{1}{\tau}\sim\frac{1}{r_{0}}\label{eq:R_const_end}
\end{equation}

For our present purposes, we can define a Capillary number $Ca$ by
comparing the imposed strain rate, $\dot{\gamma}$ for simple shear,
with the above relaxation time $Ca\equiv\dot{\gamma}\tau$. For $Ca\ll1$,
droplets will be mostly spherical, for $Ca\sim1$ droplets will be
appreciably deformed by flow, and for $Ca\gg1$ droplets will break
apart in flow.

The ideas outlined to this point - an ellipsoidal droplet deforming
non-affinely in flow and relaxing in configuration-space along a manifold
of conserved volume - are not new \cite{maffettone1998equation, minale2010models, mwasame2017macroscopic},
but it is important to re-develop these foundations before introducing
new ideas in the section that follows.

\subsubsection{\label{subsec:shape_averaging}Shape Averaging and the Log Transform}

In the preceding sections, we considered deformation and relaxation
for a single ellipsoidal droplet with shape defined by a tensor \textbf{$\boldsymbol{C}$}.
Here, we will move towards equations that extend these ideas towards
a population of droplets with different shapes and sizes, including
consideration for population balances and the effects of droplet breakup.

First, we re-interpret $\boldsymbol{C}(v)$ as a representative average
ellipsoidal droplet shape for a droplet of size $v$. Assuming an
underlying population of $N$ droplets with distinct shapes $\boldsymbol{C}_{1},\boldsymbol{C}_{2},\boldsymbol{C}_{3},...,\boldsymbol{C}_{N}$
and identical volume $v$, what is a reasonable means of computing
a representative average droplet shape? If we require that the volume
$\bar{v}$ defined by $\boldsymbol{C}(v)$ must be equal to the droplet
volume in the constitutent population, then we cannot define $\boldsymbol{C}$
as an arithmetic mean. Instead, we must pursue a tensorial analogue
of the geometric mean:

\begin{equation}
\ln(\boldsymbol{C})=\frac{1}{N}\sum_{i=1}^{N}\ln(\boldsymbol{C}_{i})\label{eq:log_average_def}
\end{equation}

taking the trace of equation \ref{eq:log_average_def}, it is now
trivial to show that if $\det(\boldsymbol{C}_{i})=1$ then $\det(\boldsymbol{C})=1$
and our representative dropet shape has the correct volume. Since
the log-transform provides a natural space for shape averaging operations,
we will re-cast our shape evolution equation in terms of $\Tb(v)=\ln(\boldsymbol{C}(v))$.
First, considering only the effects of deformation:

\begin{equation}
\ppt\Tb(v)+\ub\cdot\nabla\Tb(v)=\Tb(v)\cdot\boldsymbol{\Omega}^{T}(v)+\boldsymbol{\Omega}(v)\cdot\boldsymbol{\Tb}(v)+\boldsymbol{B}(v)+\cdots\label{eq:log_deformation}
\end{equation}

where $\boldsymbol{\Omega}$(v) and $\boldsymbol{B}$(v) are shape-dependent
projections of the rotational and extensional components of the deformation,
distinct from the usual symmetric and anti-symmetric portions of the
strain rate $\nabla\ub$. The details of this log-transformation,
including the method of calculaiton for tensor $\boldsymbol{B}$ and
$\boldsymbol{\Omega}$ has been worked out elsewhere for the upper
convected Maxwell derivative \cite{fattal2004constitutive}, and the
generalization to Gordon Schowalter simply requires substituting the
non-affine velocity gradient tensor $\kz$ for the affine one $\boldsymbol{\kappa}$.
Interestingly, however, our use of the log-transform is the first
to invoke a physical as opposed to numerical argument for it usage.
Elsewhere, the log-transformation is widely used for its advantage
in numerical stability for CFD calculations of viscoelastic fluids
\cite{fattal2004constitutive,hulsen2005flow,fattal2005time,afonso2009log}.

In equation \ref{eq:log_deformation}, the reader should note that
$\boldsymbol{\Omega}$ and $\boldsymbol{B}$ are themselves functions
of the droplet shape $\Tb$, so equations \ref{eq:log_average_def}
and \ref{eq:log_deformation} are not entirely self-consistent: the
deformation of a log-average droplet shape is now distinct from the
log-average of the individual deformed droplet shapes. Here and elsewhere,
however, we consider these kinds of pre-averaging approximations as
sub-dominant to more significant sources of uncertainty in the model
design.

Continuing on to the log-projection of the stress relaxation term,
we write:

\begin{equation}
\ppt\Tb=\cdots+\boldsymbol{R}_{\Theta}(\Tb)
\end{equation}

where the log-projection of the relaxation tensor $\boldsymbol{R}_{\Theta}(\Tb)$
is defined in relation to the relaxation tensor $\boldsymbol{R}(\boldsymbol{C})$
in equation \ref{eq:adding_relaxation}by:

\begin{equation}
\boldsymbol{R}_{\Theta}(\Tb)=e^{-\Tb}\boldsymbol{R}(e^{\Tb})
\end{equation}

Under the log transformation, the conserved volume constraint is simply
represented as $\text{tr}(\Tb)=\boldsymbol{0}$, and the equilibrium
shape is $\Tb=\boldsymbol{0}$. Conceptually, we therefore find it
easier to formulate shape relaxation in terms of $\boldsymbol{R}_{\Theta}$
to avoid the less intuitive constraint of equation \ref{eq:R_const_begin}.

Finally, we proceed to the population balance terms. The total number
of droplets in a small control volume $\delta V$ with droplet size
in the interval $v\in[v,v+\delta v]$ is given by $n(v)\delta v\delta V$,
and the sum over individual shape tensors within that grouping is
given by $n(v)\Tb(v)\delta v\delta V$. This sum can change over time
through advection, deformation, and relaxation as before, but also
via the influx/outflux of droplets due to breakup. In the balance
equation that follows, we omit the factor of $\delta v\delta V$ from
all terms. First, we consider terms relating to advection, deformation,
and relaxation:

\begin{equation}
\ppt\big(n(v)\Tb(v)\big)+\ub\cdot\nabla(n(v)\Tb(v))=n(v)\big[\Tb(v)\cdot\boldsymbol{\Omega}^{T}(v)+\boldsymbol{\Omega}(v)\cdot\boldsymbol{\Tb}(v)+\boldsymbol{B}(v)+\boldsymbol{R}_{\Theta}(\Tb)\big]+\cdots\label{eq:shape_PBE3}
\end{equation}

Next, we continue to terms pertaining to an influx and outflux of
droplets due to breakup. In the most general case, recall from the
discussion on equations \ref{eq:ndot_B_out} and \ref{eq:ndot_B_in}
that the effects of breakup appear in the number density equation
as:

\begin{equation}
\ppt n(v)=\cdots-g(v)n(v)+\int_{v}^{\infty}dv'\beta(v,v')g(v')n(v')\label{eq:ndot_brk_general_pre}
\end{equation}

In the shape evolution equation, every time a droplet departs from
or arrives into our reference size interval, its shape is added to
or removed fromthe calculation of an average shape. Hence, in the
shape evolution equation we continue from equation \ref{eq:shape_PBE3}
and add:

\begin{equation}
\ppt\big(n(v)\Tb(v)\big)=\cdots-g(v)n(v)\Tb(v)+\int_{v}^{\infty}dv'\beta(v,v')\Tb^{(D)}(v,v')g(v')n(v')\label{eq:shape_PBE4}
\end{equation}

where $\Tb^{(D)}(v,v')$ describes the typical shape of a droplet
with volume $v$ that forms upon breakup of a droplet with size $v'>v$.
Rules for specifying the shape of daughter droplets will be discussed
in section \ref{sec:Details-of-the}. Expanding the time derivative
and substituting equation \ref{eq:ndot_brk_general_pre} for $\partial_{t}n(v)$
and assuming $n(v)>0$ we can simplify. The effects of advection,
deformation, and relaxation are given by:

\begin{equation}
\ppt\Tb(v)+\ub\cdot\nabla\Tb=\Tb(v)\cdot\boldsymbol{\Omega}^{T}(v)+\boldsymbol{\Omega}(v)\cdot\boldsymbol{\Tb}(v)+\boldsymbol{B}(v)+\boldsymbol{R}_{\Theta}(\Tb)+\cdots\label{eq:shape_PBE1}
\end{equation}

and the population balance terms for droplet breakup enter as:

\begin{equation}
\ppt\Tb(v)=\cdots+\int_{v}^{\infty}dv'g(v')\frac{n(v')}{n(v)}\beta(v,v')\Big[\Tb^{(D)}(v,v')-\Tb(v)\Big]\label{eq:shape_PBE2}
\end{equation}

Since we are primarily interested in the simplest model of breakup,
$\beta(v,v')=\delta(v-v'/2)$, equation \ref{eq:shape_PBE2} becomes:

\begin{equation}
\ppt\Tb(v)=\cdots+2g(2v)\frac{n(2v)}{n(v)}\Big[\Tb^{(D)}(v')-\Tb(v)\Big]
\end{equation}

Thus we see that an influx of daughter droplets shifts the typical
droplet configuration towards that of the incoming daughter droplet
population. This is the crucial mechanism by which our framework allows
complex daughter droplet distributions to arise from a simple binary
breakage rule - the parent droplet's state of deformation is not lost
at the moment of breakup but instead transferred to the daughter droplets
and distributed acros that droplet size interval.

Similar population balance terms can be incoporated into the constitutive
equation where there is interest in nucleation, growth, coalescence,
etc., but these are beyond the scope of interest for our initial study.

As a final note to this section, the ``trick'' of transforming to
the log-configuration $\Tb$ is sensible when the droplets are described
in strictly ellipsoidal terms. For a more general case in which droplets
are described by a collection of higher-order shape tensors, it is
our understanding that this trick is not transferrable. For higher-order
tensor genereralizations, volume conservation could be handled via
an additional isotropic term in the shape evolution equation, modulated
by a Lagrange multiplier that constrains shape evolution to a pre-defined
manifold of constant volume. For the time being, however, it is our
view that there are major technical obstacles to a higher-order tensor
implementation (e.g. computing droplet volume and surface area becomes
non-trivial) and we find the log-transform approach sufficiently elegant
to warrant a departure from a fully future-proof mathematical structure.

\section{\label{sec:Details-of-the}Details of the Constitutive Equation}

In section \ref{sec:Governing-Equations}, we worked out a general
framework for modeling dense emulsions including population balance
equations, droplet shape evolution equations, and momentum balance
equations. To this point, however, we have omitted implementation-specific
details of the couplings between these equations: how does the droplet
shape tensor $\Tb$ enter into the capillary stress $\boldsymbol{\sigma}$,
the breaking rate $g(v)$, and the shape relaxation tensor $\boldsymbol{R}_{\Theta}$?
And how does the shape of the parent droplet $\Tb(v)$ inform the
shape of the daughter droplets $\Tb^{(D)}(v)$ following breakup?
In this section, we will sketch a strategy that resolves these questions,
but we cannot guarantee the performance of our scheme under any specific
engineering application: the details of this section may be subject
to revision and improvement as suitable experimental/modeling data
is available for confrontation.

To close the STEPB model, we first define a droplet free energy in
terms of an estimated surface area. Next, we can describe the associated
droplet stress and propose expressions for shape relaxation, checking
for positive entropy production and conservation of droplet volume.
Finally, we will define daughter droplet shapes and kinetic expressions
for droplet breakup.

\subsection{Droplet Surface Area Estimation}

We will assume that the free energy $F$ of an isolated ellipsoidal
droplet is just its surface area, weighted by a surface tension $\Gamma$.
The surface tension $\Gamma$ will be assumed to remain constant in
time and uniform across the droplet surface, both of which are possible
in applications where there is a large excess of surfactant present
in solution (e.g. mayonnaise and creams).  Future iterations on this
modeling framework would be needed to cover systems in which the distribution
of surfactant becomes non-trivial.

While there is no closed-form expression for the surface area of an
ellipsoid, a suitable estimate - accurate to within about 1\% - can
be obtained via the Knud-Thomsen approximation \cite{xu2009ellipsoidal,mwasame2017macroscopic}:

\begin{equation}
F(\boldsymbol{C},r_{0})=4\pi r_{0}^{2}\Gamma\Big(\frac{1}{3}I_{2}(\boldsymbol{C}^{p/2})\Big)^{1/p}\label{eq:droplet_F}
\end{equation}

\begin{equation}
I_{2}(\boldsymbol{X})=\frac{1}{2}(\text{tr}(\boldsymbol{X})^{2}-\text{tr}(\boldsymbol{X}^{2}))
\end{equation}

where $I_{2}$ is the second frame invariant and the value $p=1.6$
optimizes the approximation across a wide space of ellipsoids. A similar
estimation of surface area was previously made for highly faceted
emulsions, where $p=2$ was considered more appropriate \cite{larson1997elastic}.
We will continue to use $p=1.6$ for the remainder of this report
to remain internally consistent with our ellipsoidal approximation
of droplet shapes, but within this range we do not consider the choice
of $p$ to be extremely important compared to other assumptions and
approximations within the model.

To compute the total free energy $F_{T}$ of a collection of droplets
with log-averaged shape tensor $\Tb$, we rely on a pre-averaging
approximation - within a given volume range $v\in[v,v+\delta v]$,
we assume that the surface area of the average droplet shape is the
same as the average surface area of the underlying droplet shapes.
Summing over the free energy from all possible droplet sizes, we obtain:

\begin{equation}
F_{T}=\int_{0}^{\infty}dvn(v)F(e^{\Tb(v)},r_{0}(v))
\end{equation}

This pre-averaging approximation works best when the droplet shape
at any size $v$ is narrowly distributed. For more broadly dispersed
droplet shape distributions, our pre-averaging approximation will
underestimate the free energy. Unfortunately, in our view it is not
possible to improve on this preaveraging approximation without compromising
the overall simplicity and computational cost of the framework outlined
in section \ref{sec:Governing-Equations}.

\subsection{Droplet Stress}

The reversible work needed to deform a droplet must be equal to the
change in free energy associated with the change in droplet shape
\cite{larson2013constitutive, schieber2021nonequilibrium}. Since
we know that the shape tensor $\boldsymbol{C}$ is deformed by a Gordon-Schowalter
derivative and the free energy in equation \ref{eq:droplet_F} is
defined explicitly in terms of that same shape tensor, it is possible
to define the droplet's elastic stress $\boldsymbol{\sigma}(\boldsymbol{C})$
by which the fluid resists deformation \cite{larson2013constitutive}:

\begin{equation}
\boldsymbol{\sigma}(\boldsymbol{C},v)=2(1-\zeta)\boldsymbol{C}\cdot\frac{\delta F}{\delta\boldsymbol{C}}
\end{equation}

Evaluating the derivative on $F$ using equation \ref{eq:droplet_F},
this becomes:

\begin{equation}
\boldsymbol{\sigma}(\boldsymbol{C},v)=n(v)v\Big(\frac{\Gamma}{r_{0}(v)}\Big)(1-\zeta)(I_{2}(\boldsymbol{C}^{p/2})^{(1/p-1)}(\text{tr}(\boldsymbol{C}^{p/2})\boldsymbol{C}^{p/2}-\boldsymbol{C}^{p})\label{eq:single_droplet_stress}
\end{equation}

Note that the capillary stresses are proportional to a modulus $G_{0}=\Gamma/r_{0}$
that increases with surface tension and decreases with droplet radius.
To compute the total stress in a collection of droplets, we once again
use our pre-averaging approximation and assume that the average stress
from a collection of equally-sized droplets is the same as the stress
from a droplet with average shape:

\begin{equation}
\boldsymbol{\sigma}_{T}=\int_{0}^{\infty}dv\boldsymbol{\sigma}(e^{\Tb(v)},r_{0}(v))
\end{equation}

\subsection{Shape Relaxation}

As discussed in section \ref{subsec:Constraints-on-shape}, the shape
relaxation tensor $\boldsymbol{R}_{\Theta}(\Tb)$ must be traceless
to conserve droplet volume, $\text{tr}(\boldsymbol{R}_{\Theta})=0$,
and it must also ensure positive entropy production to be thermodynamically
consistent. The simplest means of satisfying these constraints is
to assume that stress relaxation is linear in $\Tb$:

\begin{equation}
\ppt\Tb(v)=\cdots-\frac{1}{\tau(v)}\Tb+\cdots
\end{equation}

where the relaxation time $\tau(v)$ depends on the droplet volume
and is found by balancing elastic and viscous forces during the shape
relaxation process, as discussed in section \ref{subsec:Constraints-on-shape}.
In particular, the relaxation time $\tau(v)$ scales inversely with
the droplet's equilibrium radius, $\tau(v)\sim v^{-1/3}$. Projecting
this expression for shape relaxation back to the shape tensor $\boldsymbol{C}=e^{\Tb}$,
this becomes:

\begin{equation}
\ppt\boldsymbol{C}=\cdots-\frac{1}{\tau(v)}\boldsymbol{C}\cdot\ln\boldsymbol{C}
\end{equation}

In the figure below, we provide a contour plot of the per-droplet
rate of entropy production $\dot{S}$ (in units of $v\mu^{eff}/\tau^{2}(v)$)
as a function of the two independent eigenvalues, $\lambda_{1}$ and
$\lambda_{2}$ of a configuration tensor $\boldsymbol{C}$. Note that
the third eigenvalue $\lambda_{3}$ is constrained by $\det(\boldsymbol{C})=\lambda_{1}\lambda_{2}\lambda_{3}=1$.

\begin{equation}
\dot{S}(\boldsymbol{C})=-\frac{\partial F}{\partial\boldsymbol{C}}:\boldsymbol{R}(\boldsymbol{C})=\frac{1}{\tau}\frac{\partial F}{\partial\boldsymbol{C}}:(\boldsymbol{C}\cdot\ln(\boldsymbol{C}))\label{eq:Sdot_def}
\end{equation}

\begin{figure}
\begin{centering}
\includegraphics[scale=0.6]{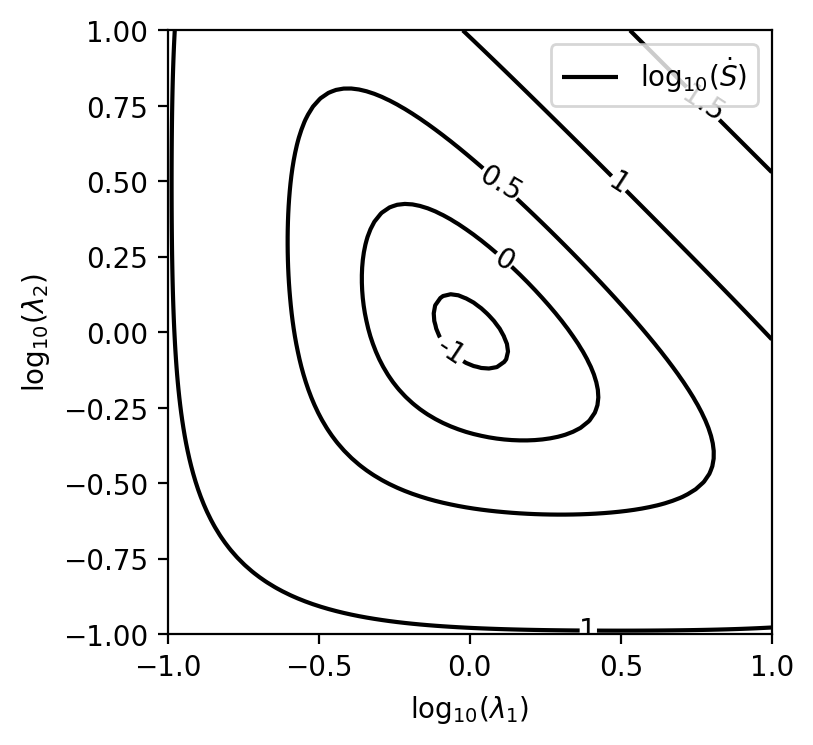}
\par\end{centering}
\caption{\label{fig:entropy_production}Demonstrating positive entropy production
per equation \ref{eq:Sdot_def} over a wide range of possible droplet
shapes, as defined in terms of the two independent eigenvalues $\lambda_{1}$
and $\lambda_{2}$ of the droplet shape tensor $\boldsymbol{C}$.
Note that we plot contours of $\log_{10}(\dot{S})$ for more even
spacing, so contours with negative labels still represent positive
entropy production. Per-particle entropy production has assumed units
of $v\mu^{eff}/\tau^{2}(v)$ in this figure.}
\end{figure}

In figure \ref{fig:entropy_production}, we see that entropy production
goes to zero when the droplet shape is at equlibrium, $\boldsymbol{C}=\boldsymbol{I}$,
and it is strictly positive for droplets deformed from equilibrium.
We have also explored a wider range of simple expressions for $\boldsymbol{R}(\boldsymbol{C})$
and it does not appear to be difficult to construct relaxation tensors
that satisfy the constraints of equations \ref{eq:R_const_begin}
- \ref{eq:R_const_end}. In our experience, any reasonably constructed
relaxation tensor that conserves droplet volume (the easier constraint
to design around) tends to also show positive entropy production for
the free energy as defined in equation \ref{eq:droplet_F}.

To close this section, we note that a considerable body of work already
exists for describing the shape evolution of ellipsoidal droplets
in \emph{dilute emulsions}, complete with comparisons to experimental
data \cite{maffettone1998equation, minale2010models, mwasame2017macroscopic}.
One might reasonably argue that our model could be immediately improved
by applying the tried-and-tested shape relaxation tensors employed
therein, but we defer such an approach for the time being for two
reasons: (1) what is true about shape relaxation in dilute emulsions
may not be true for dense emulsions, since the limits of our ``effective
medium approximation'' are not yet established from experiments,
and (2) the shape relaxation tensor is not necessarily the main deficiency
of the overall modeling framework, and corrections on $\boldsymbol{R}_{\Theta}$
will make the overall modeling framework more complex but not necessarily
better given the limitations that persist elsewhere.

\subsection{Daughter Droplet Shapes\label{subsec:Daughter-Droplet-Shapes}}

The conditions under which droplets can break apart will be covered
in section \ref{subsec:Breakup-Dynamics}, but we will briefly preempt
that discussion to think about the shape of the daughter droplets
left behind after droplet breakup takes place. In keeping with the
framework outlined in section \ref{subsec:shape_averaging}, we will
assume that daughter droplets are purely ellipsoidal in their shape.

Here, we consider the following scenario: prior to breaking, a parent
droplet with volume $v$ has shape tensors $\boldsymbol{C}$ and $\Tb=\ln(\boldsymbol{C})$,
and the breakup process produces two identical daughter droplets with
volumes $v/2$ and shape tensors $\boldsymbol{C}^{(D)}$ and $\Tb^{(D)}=\ln(\boldsymbol{C}^{(D)})$.
Ideally, we would like to specify the droplet shape such that there
is perfect continuity across the breaking event - the daughter droplet
would have the same surface area and the same stress as the parent
droplets, for example. Unfortunately, however, it is not possible
to ensure continuity across both these measures.

An ellipsoid is specified through six independent pieces of information:
three to define the length of each axis, and three more to orient
it in space. If we assign the orientation of the daughter droplet
to that of the parent droplet, then we need three additional constraints
to specify the daughter droplet shape. This can be achieved by matching
the principle components of the parent/daughter droplet's deviatoric
stress tensors:

\begin{equation}
\boldsymbol{\sigma}(\boldsymbol{C}^{(D)},v/2)+\Lambda I=\frac{1}{2}\boldsymbol{\sigma}(\boldsymbol{C},v)
\end{equation}

Specifying the daughter droplet shape in this way, it is no longer
possible to constrain the change in surface area that occurs through
breaking. In fact, we find that conserving droplet stress always leads
to an increase in droplet surface area following breakup, in violation
of basic thermodynamic principles. Conversely, if we constrain the
daughter droplet shapes to a manifold of specified total surface area,
then in our view there is no clear basis to specify three axis of
an ellipsoid with just two additional pieces of information.

At this point, it is helpful to remember that droplet breakup is,
in reality, defined by a discrete topological transition (from one
closed manifold to two or more) that takes place in the course of
droplet shape evolution. With a higher-order tensor representation
of the droplet shape, it would be possible to resolve such a transition
explicitly, with all the relevant physics encoded in shape relaxation
tensors like $\boldsymbol{R}(\boldsymbol{C})$. Having constrained
ourselves to ellipsoidal projections of droplet shape, however, we
have lost the capacity to resolve topological transitions explicitly,
and instead must resolve them implicitly. Thus, we argue that inquiries
on thermodynamic consistency must treat shape relaxation and droplet
breakup as joint processes, assessing only the net entropy production
of the two processes combined. In section \ref{subsec:Large-strains-with}
we will show that if the kinetics for breakup and shape relaxation
are bound to similar timescales, the net entropy production from shape
relaxation will indeed be positive.

To close this section, some readers may find our choice to conserve
droplet stress across breakup surprising: after all, surely there
must be some stress relief associated with the breakup process? Here,
we remind the reader that we are effectively using the daughter droplets
and parent droplets together as a means of interpolating properties
of the intermediate states that lead to breakup - some portion of
stress relaxation in the daughter droplet distribution is effectively
serving as a proxy for fast-relaxing capillary modes that would exist
in a higher-order representation of the parent droplet shape. Therefore,
conserving stress at the moment of breakup does not preclude an indirect
benefit to stress relaxation across the whole breakup process. A similar
coupling between stress relaxation and breakup has previously been
described for polymeric materials \cite{cates1987reptation,peterson2020full}. 

\begin{figure}
\begin{centering}
\includegraphics[scale=0.6]{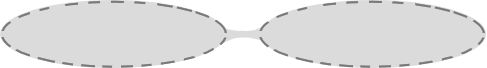}
\par\end{centering}
\caption{\label{fig:droplet_cartoon}Cartoon representation for our interpretation
of the ellipsoidal droplet approximation as it pertains to assigning
daughter droplet shapes. The above droplet is technically unbroken
- topologically speaking it is a single non-ellipsoidal manifold,
but many of its physical properties (stress, surface area, etc) can
be reasonably captured via a pair of ellipsoids in this case. Daughter
droplet shapes are defined such that stress relaxes continuously through
the whole breakup process. }
\end{figure}

This same idea can be communicated graphically in the cartoon of Figure
\ref{fig:droplet_cartoon}: topologically speaking there is only one
droplet, but in many respects its shape is better described as two
ellipsoids arranged head-to-tail. To choose the pair of ellipses that
best captures the contorted shape of the parent droplet, we suggest
 matching the capillary stresses. This assumption is physically grounded
in the overall continuity of the breakup process - the stress relaxation
attributed to breakup is realized through a gradual shape relaxation
and not an instantaneous topological transformation.

\subsection{Breakup Dynamics\label{subsec:Breakup-Dynamics}}

In dilute emulsions, droplet breakup is deterministic based on the
bulk strain history. For dense suspensions, however, deformation at
the scale of individual droplets is less predictable - for the same
bulk deformation, individual droplets can experience different deformations
and thus breakup times become more broadly distributed.

Working from our ellipsoidal approximation of droplet shape, there
are two natural ways that one might represent the breaking rate $g(v)$
in equation \ref{eq:ndot_final}. First, as might be appropriate for
dilute emulsions, we could say that there exists a bounding manifold
in shape-space $f(\Tb)=0$ that defines a ``breaking surface''.
Droplets within the envelope of this manifold remain cohesive but
split spontaneously upon reaching the breaking surface leading to
a breaking rate that is delta-distributed in time:

\begin{equation}
g(v)=\delta(f(\Tb(v)))
\end{equation}

Even for dilute emulsions, however, this expression is problematic
when the shape is constrained to be ellipsoidal. For higher-order
descriptions of the droplet shape, the key property of the breaking
surface is that it marks the boundary of a topological transition,
and no such boundary can be marked on the space of ellipsoids.

For dense emulsions, delta-distributed breaking rates are also unacceptable
on a more fundamental level - one would prefer a distribution of breaking
times to reflect the distribution of droplet-scale strain histories.
The simplest possible approximation is to assume that for a given
average droplet shape $\Tb$, breaking times are Poisson-distributed
(memoryless) with a driving force towards breakup that can be inferred
from the average shape alone via some function $h(\Tb,v)$. We further
assume that the timescale for breakup is tied to the timescale for
shape relaxation, with some $O(1)$ proportionality factor $\alpha$:

\begin{equation}
g(v)=\frac{\alpha}{\tau}h(\Tb,v)
\end{equation}

Following our discussion in section \ref{subsec:Daughter-Droplet-Shapes},
we note that the prefactor $\alpha$ will have an important role to
play with respect to guaranteeing the thermodynamic consistency of
the joint breaking/shape relaxation process. For simplicity, we will
assume that the driving force for breakup is proportional to the difference
in surface area between the current state (deformed ellipsoidal droplet)
and a hypothetical end-state of the shape relaxation/breakup process
(two spherical droplets):

\begin{equation}
F_{brk}(r_{0}(v))=2F(\boldsymbol{I},r_{0}(v/2))
\end{equation}

\begin{equation}
h(\Tb,v)=\begin{cases}
0 & \text{if }F(e^{\Tb},r_{0}(v))\leq F_{brk}(r_{0}(v))\\
F(e^{\Tb},r_{0}(v))/F_{brk}(r_{0}(v))-1 & \text{if }F(e^{\Tb},r_{0}(v))>F_{brk}(r_{0}(v))
\end{cases}
\end{equation}

Recall again that the shape tensors $\boldsymbol{C}$ and $\Tb$ are
related by $\boldsymbol{C}=e^{\Tb},$and $F(\boldsymbol{C},r_{0})$
is the free energy (proportional to surface area) of a droplet with
equilibrium radius $r_{0}$ and ellipsoidal shape defined by $\boldsymbol{C}$.
This simplistic picture of $h(\Tb,v)$ qualitatively captures an increasing
proclivity to breakup with increasing extent of deformation, but it
has at least one major weakness in that it takes no consideration
for the type of deformation being applied. For example, the same change
in surface area could be achieved with either a prolate or oblate
spheroid, but the former generally is far more susceptible to breakup.

More general rate expressions (e.g. including memory functions) are
possible but in our view difficult to parameterize in a simple but
compelling way. As such, we feel that a Poisson-approximation of breaking
time distributions provides the most appropriate balance of simplicity
and insight for our present interests.

This concludes our presentation for the governing equations of the
STEPB model.

\section{\label{sec:step-shear}A Path to Parameterization: Sample Calculations
for Monodisperse Systems}

Having finalized the details of STEPB in section \ref{sec:Details-of-the},
here we will provide a first set of calculations and discuss a path
towards parameterizing the model and testing its predictions on the
basis of comparisons to experimental data. As mentioned in the introduction,
for the purposes of quantitatively testing our model there is presently
an insufficient body of experimental work on time-dependent droplet
breakup in stabilized dense emulsions. Therefore, we will instead
use this opportunity to propose a new design of experiments that could
challenge the predictions of STEPB and determine its usefulness as
a framework for modeling dense emulsions. 

Droplet breakup in monodisperse dense emulsions following a step shear
deformation poses a number of conceptual challenges for existing models
\cite{dubbelboer2016pilot, maindarkar2014prediction, doi1991dynamics}.
In step shear, the processes of deformation and relaxation/breakup
do not occur concurrently but sequentially. This separation of deformation
and breakup is important for describing processes at high Deborah
number, where the Deborah number $De$ compares the timescale for
\emph{changes in strain rate} against the droplet's natural relaxation
time $\tau$. Only viscoelastic models are capable of appropriately
separating deformation and breakup, and to our knowledge the STEPB
model is the first to include the requisite  coupling between the
population balance equations and the droplet shape evolution equations. 

In the absence of exerimental data, we will briefly discuss the an
expected outcomes of the proposed step-deformation experiment - this
will guide our interpretation of the model predictions that follow.
First, we expect that for small deformations, droplets will relax
back to an isotropic shape without breaking, leading to no net change
in the droplet size distribution. Above some critical deformation,
a fraction (not all) of the droplets will begin to break. This is
distinct from what should be expected in dilute emulsions, where outcomes
are strictly determined. Finally, with increasing initial deformation
more and more of the droplets will break into smaller and smaller
fragments. These expected outcomes seem reasonable, but they should
not stand unchallenged - it is our view that studying the dynamics
of droplet breakup in dense emulsions should be just as productive
today as studying the dynamics of droplet breakup in dilute emulsions
was four decades ago.

As a further motivation to consider step-shear calculations, step-shear
provides a self-contained parameterization of STEPB where complete
experimental data is available. As currently formulated, the STEPB
model can be fully specified through an initial droplet size distribution
plus four additional parameters, namely the surface tension $\Gamma$,
a slip parameter $\zeta$, a reference droplet relaxation time $\tau$,
and a kinetic prefactor $\alpha$ governing droplet breakup dynamics.
The viscosity $\mu^{eff}$ is omitted from this list, since it obviated
by the assumed relation $\mu^{eff}\sim\Gamma\tau/r_{0}$. Our parameter
list also lacks explicit reference to the viscosity ratio of the component
Newtonian fluids - the viscosity ratio should end up modulating the
slip parameter $\zeta$, as is extensively documented  in the literature
on dilue emulsions \cite{minale2010models}. The viscosity ratio should
also modulate other model parameters, including the kinetic prefactor
$\alpha$ and the precise relationship between $\mu^{eff}$ and $\tau$.
At present, there is insufficient data on droplet relaxation/breakup
in dense emulsions to properly codify an explicit dependence on the
viscosity ratio, so these details must be deferred to future work.

In section \ref{subsec:Nonlinear-Elasticity:-Modulus}, we will show
that the surface tension $\Gamma$ and slip parameter $\zeta$ can
be found through the linear and nonlinear elasticity of an emulsion.
In section \ref{subsec:Medium-strains-without}, we will show that
a relaxation time $\tau$ is evident through the stress relaxation
dynamics. In section \ref{subsec:Large-strains-with}, we will show
that the kinetic prefactor $\alpha$ changes both the final particle
size distribution and the stress relaxation kinetics. Finally, in
section \ref{subsec:Strain-Sweep-to} we will explore trends for strain-dependent
daughter droplet distributions that emerge in the STEPB model.

The calculations that follow consider a monodisperse dense emulsion
with droplet volume $v_{0}$ and total volume fraction $\phi$. Note
that the total volume fraction is related to the number density distribution
by $\phi=\int_{0}^{\infty}n(v)vdv$, so that $n(v)=\delta(v-v_{0})\phi/v_{0}$
for the case of monodisperse emulsions.

For additional details on the numerical methods of solution, we refer
the reader to appendix \ref{sec:Addition-Details-on}.

\subsection{\label{subsec:Nonlinear-Elasticity:-Modulus}Nonlinear Elasticity:
Modulus and 'slip' parameter}

The surface tension of a monodisperse dense suspension can be indirectly
measured through its linear elastic response to deformation, and the
propensity to 'slip' can be similarly inferred from a nonlinear elastic
response. Here, we will consider predictions for the linear and nonlinear
elasticity with respect to step shear, since shear deformations are
typically easier to realize in experiments.

\begin{figure}
\begin{centering}
\includegraphics[scale=0.6]{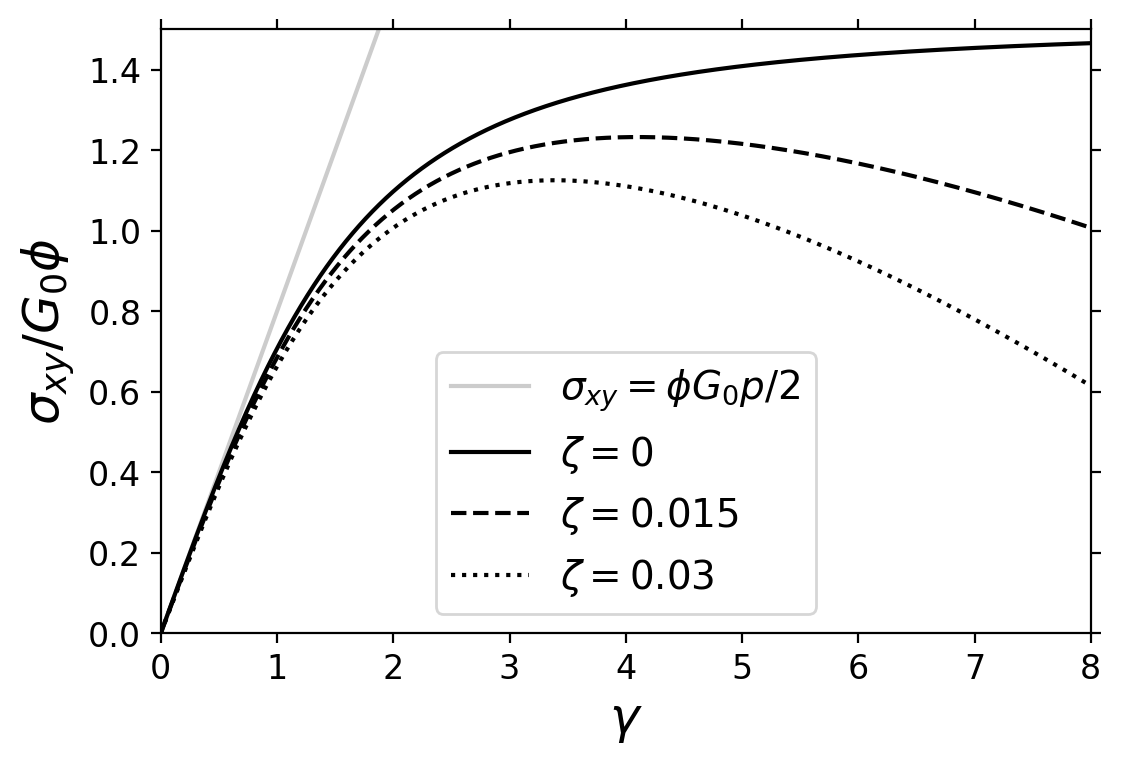}
\par\end{centering}
\caption{\label{fig:Nonlinear_Elasticity}STEPB model redictions for the linear
and nonlinear elasticity of a monodisperse dense emulsion under step
shear deformation. The linear elastic response $\sigma_{xy}=\phi G_{0}p/2$
provides a means of estimating the surface tension via $G_{0}=\Gamma/r_{0}$.
The nonlinear elastic response provides a means of further fine-tuning
the model to account for a slip parameter $\zeta$ as and when such
corrections are necessary.}
\end{figure}

In Figure \ref{fig:Nonlinear_Elasticity}, we plot the dimensionless
shear stress $\sigma_{xy}/\phi G_{0}$ as a function of the shear
strain amplitude $\gamma$. First, we note that the linear elastic
response has a modulus $G=G_{0}p/2$, where $p=1.6$ in the Knud-Thomsen
approximation. Strain-softening in the absence of slip occurs at high
strain where changes in droplet surface area become linear (as opposed
to quadratic) in the applied strain. Inclusion of a slip parameter
$\zeta$ enhances strain softening in the nonlinear elastic response,
as droplets begin rotating away from the axis of extension without
stretching. At sufficiently large strains, even small amounts of 'slip'
can lead to dramatic changes and 'tumbling' when droplets over-rotate
and shear stress becomes an oscillatory function of strain amplitude.

Where experimental data is available, the linear elastic response
provides a means of estimating the surface tension $\Gamma$, via
$\Gamma=2Gr_{0}/p$. The nonlinear elastic response can also inform
the description of 'slip' effects wherever $\zeta=0$ under-predicts
strain softening at large strains. The choice of $\zeta$ will influence
the critical capillary number in steady shear flow (c.f. section \ref{sec:experiments}),
and ideally correcting for slip in the elastic response should provide
a more accurate prediction of the critical capillary response in steady
shear - where this is not the case, however, it must be recognized
that the way we describe droplet deformation bypasses the underlying
hydrodynamics problem and is thus subject to error. In practice, therefore,
therefore, it may often be more useful to use $\zeta$ as a means
of tuning the critical capillary number (c.f. appendix \ref{sec:Slip-and-the})
rather than the nonlinear elastic response to step deformation.

\begin{figure}
\begin{centering}
\includegraphics[scale=0.6]{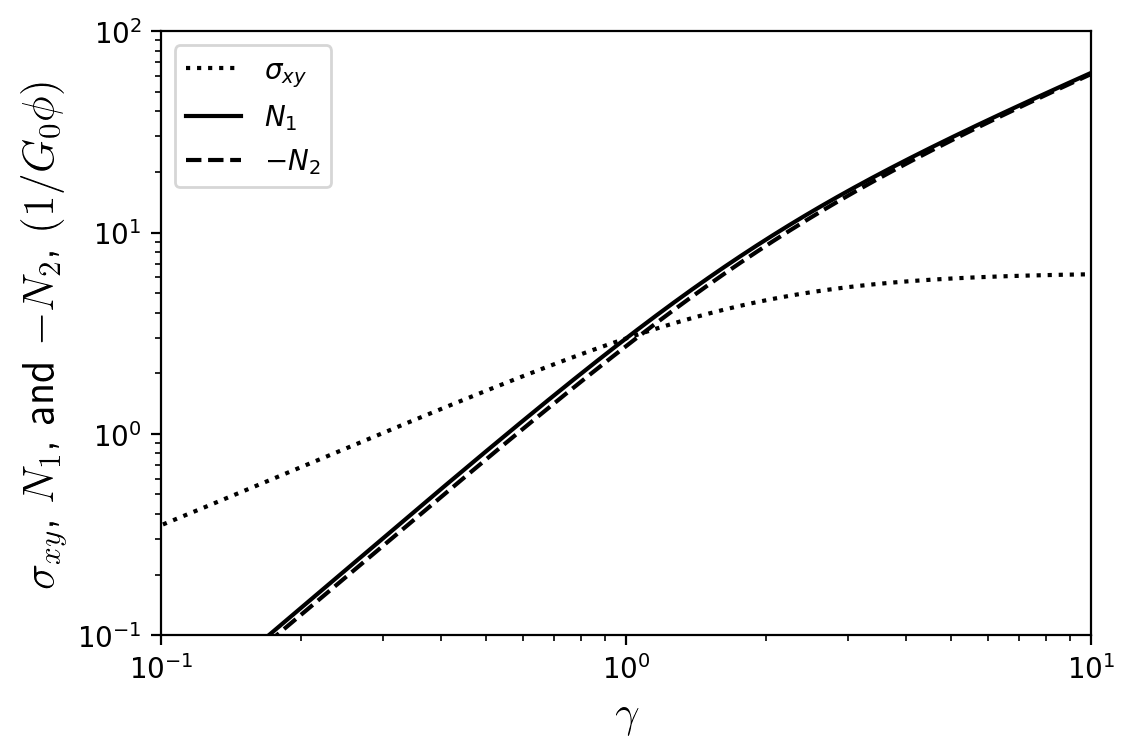}
\par\end{centering}
\caption{\label{fig:step_shear_N1N2}Comparing the normal stress differences
$N_{1}$ and $-N_{2}$ with the shear stress $\sigma_{xy}$ over a
range of strain amplitudes $\gamma$. Across all strain amplitudes,
the first and second normal stress differences are comparable in scale
but opposite in sign. At low strains, normal stresses increase quadratically,
while at high strains they increase linearly.}
\end{figure}

For completeness, figure \ref{fig:step_shear_N1N2} gives STEPB model
calculations of the first and second normal stress differences in
step shear for $\zeta=$0. The results are similar to models discussed
elsewhere in the literature \cite{larson1997elastic}, in that the
first and second normal stress differences are comparable in magnitude
but opposite in sign, and normal stresses increase linearly with strain
for $\gamma\gg1$.

Unless stated otherwise, all calculations that follow will assume
$\zeta=0$.

\subsection{\label{subsec:Medium-strains-without}Medium strains without breakup:
linear relaxation time}

At large strains, droplets will be prone to breakup, but small strains
in dense suspensions will often have a yield stress that precludes
observation of a shape relaxation timescale. Therefore, intermediate
strains may be most useful for parameterizing the STEPB model's shape
relaxation time $\tau$. However, because the stress tensor $\boldsymbol{\sigma}_{T}$
and relaxation tensor $R_{\Theta}$ are not linearly related to one
another, there is no a-priori guarantee that stress relaxation following
nonlinear deformations will yield a single easily measured relaxation
time. If nonlinearities dominate stress relaxation at early times
for intermediate strain deformations, then it may be difficult to
uniquely parameterize the model, since droplet breakage is a confounding
source for nonlinearity in the overall stress relaxation. In the absence
of experimental data for comparison, this section will simply test
whether such nonlinearities dominate stress relaxation in the STEPB
model, since this is a necessary (albeit not sufficient) condition
for the success of our proposed parameterization strategy.

\begin{figure}
\begin{centering}
\includegraphics[scale=0.6]{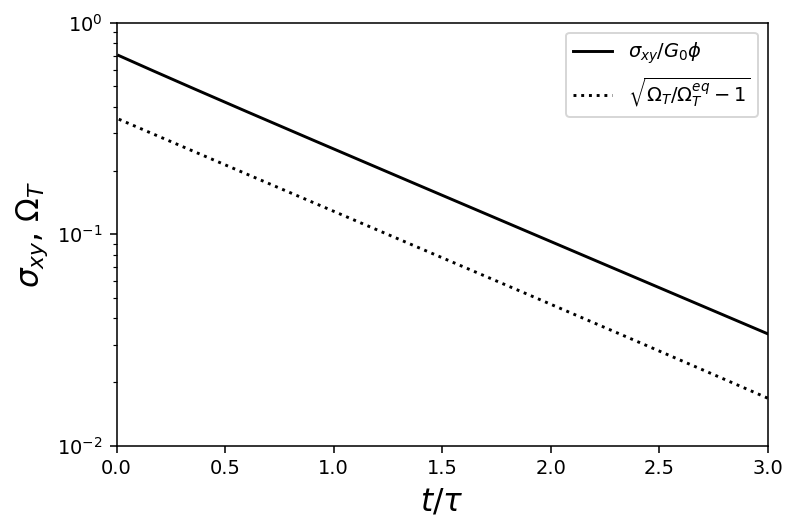}
\par\end{centering}
\caption{\label{fig:linear_relaxn}Predictions for stress relaxation following
a medium-amplitude deformation, where the capillary stresses are more
reliably measured and droplet breakup is not yet taking place. In
the STEPB model, the recovery of both stress and surface area can
be described as single-exponential processes on the time interval
considered above.}
\end{figure}

Here, we consider predictions of the STEPB model for strain $\gamma=$1.
This strain is below our proposed critical strain for breakup $\gamma=1.5$
(c.f. section \ref{subsec:Breakup-Dynamics}), but well above a typical
yield strain for weakly jammed dense emulsions, $\gamma\sim0.03$
\cite{scheffold2013linear}. The predictions are shown in figure \ref{fig:linear_relaxn},
wherein it is evident that stress relaxation is effectively linear,
with relaxation time $\tau$, over the entire time interval of interest.
Where a yield surface is present, the linear relaxation process will
be interrupted as the droplets approach a stable isostatic configuration.
Figure \ref{fig:linear_relaxn} also shows that stress relaxation
is directly correlated to a decrease in the total droplet surface
area $\Omega_{T}$ relative to the pre-strained surface area $\Omega_{T}^{eq}$.

Where the relaxation time $\tau$, the surface tension $\Gamma$,
and the monodisperse droplet size $r_{0}$ can all be measured independently,
it is possible to infer an estimate of the effective emulsion viscosity
$\mu^{eff}$ indirectly, via $\mu^{eff}=\Gamma\tau/r_{0}$. This estimate
of $\mu^{eff}$ can be further tested through steady shear measurements,
provided (1) stresses are well above the yield stress and well below
$Ca\sim1$ and (2) the viscous contribution to the stress is appropriately
isolated from the capillary stress contribution.

\subsection{Large strains with breakup: breakup kinetics\label{subsec:Large-strains-with}}

Continuing to large strains, here we explore both implicit and explicit
signatures of droplet breakup in the STEPB model. Implicit signatures
of droplet breakup can be found by observing stress relaxation, whereas
explicit signatures of droplet breakup are evident through direct
measurement of a droplet size distribution. Assuming parameters $\Gamma,\zeta,$
and $\tau$ are already known through independent means, these measurements
can help assign the kinetic prefactor to breakup, $\alpha$, which
controls the relative timescales for shape relaxation and droplet
breakup. Because droplet surface area increases discontinuously for
individual breakup events, we first need to check that our choice
of $\alpha$ maintains a thermodynamically consistent joint description
of relaxation/breakup.

\begin{figure}
\begin{centering}
\includegraphics[scale=0.6]{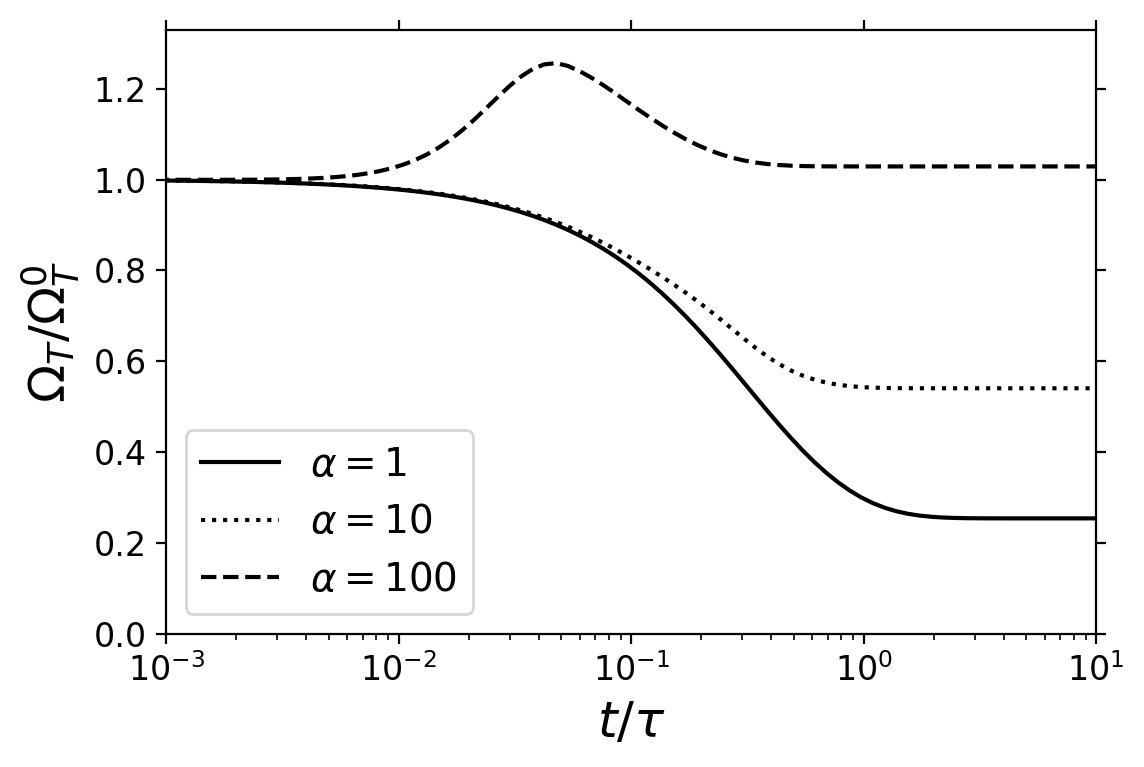}
\par\end{centering}
\caption{\label{fig:totalSA_t}Predictions of the STEPB model with $\zeta=0$
for total surface area during the process of relaxation from an initial
step strain of $\gamma=10$. The kinetic prefactor to the breaking
rate, $\alpha$, controls how quickly breaking occurs relative to
shape relaxation. Larger values of $\alpha$ allow more breaking events
to occur, and leading to a steady state with smaller average droplet
size and higher overall surface area. Excessively large values of
$\alpha$ are problematic, however, leading to a transient increase
in the total surface area. The source of this non-physical behavior,
is discussed in the main text.}
\end{figure}

Here, we evaluate the total surface area $\Omega_{T}$ normalized
by the initial surface area after deformation $\Omega_{T}^{0}$ following
a step-strain of $\gamma=10$. Predictions are shown for $\alpha=1,10,100$
in figure \ref{fig:totalSA_t}. For $\alpha=1,10$ there are no issues
- the total surface area is always decreasing in time, and increasing
$\alpha$ allows droplets to break faster and preserve more of the
extra surface area created during the initial deformation. However,
for $\alpha=100$ droplets are able to break much faster than they
can relax and since each successive break leads to an increase in
surface area (c.f. section \ref{subsec:Daughter-Droplet-Shapes})
we observe a non-physical transient increase in the total surface
area.

Because surface area must continually decrease during relaxation/breakup,
large values of $\alpha$ lead to a model that is thermodynamically
inconsistent. However, in constructing the STEPB model, $\alpha$
was never intended to take on any arbitrary value - it was always
intended as a means of modulating the relationship between the physically
connected processes of shape relaxation and droplet breakup, taking
on a value $\alpha\sim O(1)$. Absent experimental data for a direct
comparison, for convenience we will use $\alpha=10$ to amplify the
effects of breakage while still remaining safely in the space of thermodynamically
consistent predictions. This choice does not necessarily endorse $\alpha=$10
as a realistic value, and in application varying $\alpha$ will be
an important handle for matching experimental data.

\begin{figure}
\begin{centering}
\includegraphics[scale=0.6]{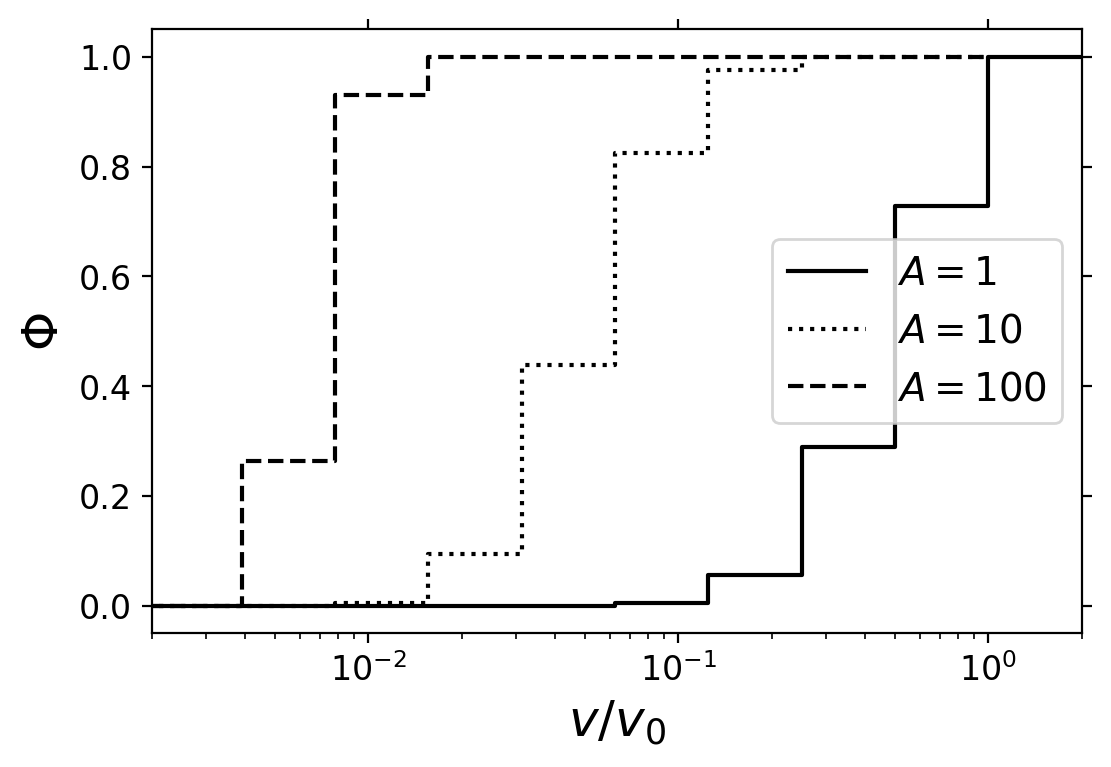}
\par\end{centering}
\caption{\label{fig:Phi_vary_alpha}The final daughter droplet size distribution
following an initial strain of $\gamma=10$ for $\alpha=1,10,100$,
$\zeta=0$. This figure provides complementary information to figure
\ref{fig:totalSA_t}.}
\end{figure}

As a complement to Figure \ref{fig:totalSA_t}, we also show in figure
\ref{fig:Phi_vary_alpha} the final droplet size distribution in terms
of a cumulative distribution function (CDF), $\Phi(v)=1/\phi\int_{0}^{v}n(v)vdv$.
Figure \ref{fig:Phi_vary_alpha} covers the same initial strain $\gamma=10$
and the same values of $\alpha=1,10,100$. With increasing $\alpha$
droplets are better able to break while they remain deformed, leading
to smaller droplet sizes in the final distribution. Note that the
CDF is a stepwise function because the droplet size distribution is
described through a series of delta-peaks in this case.

\begin{figure}
\begin{centering}
\includegraphics[scale=0.6]{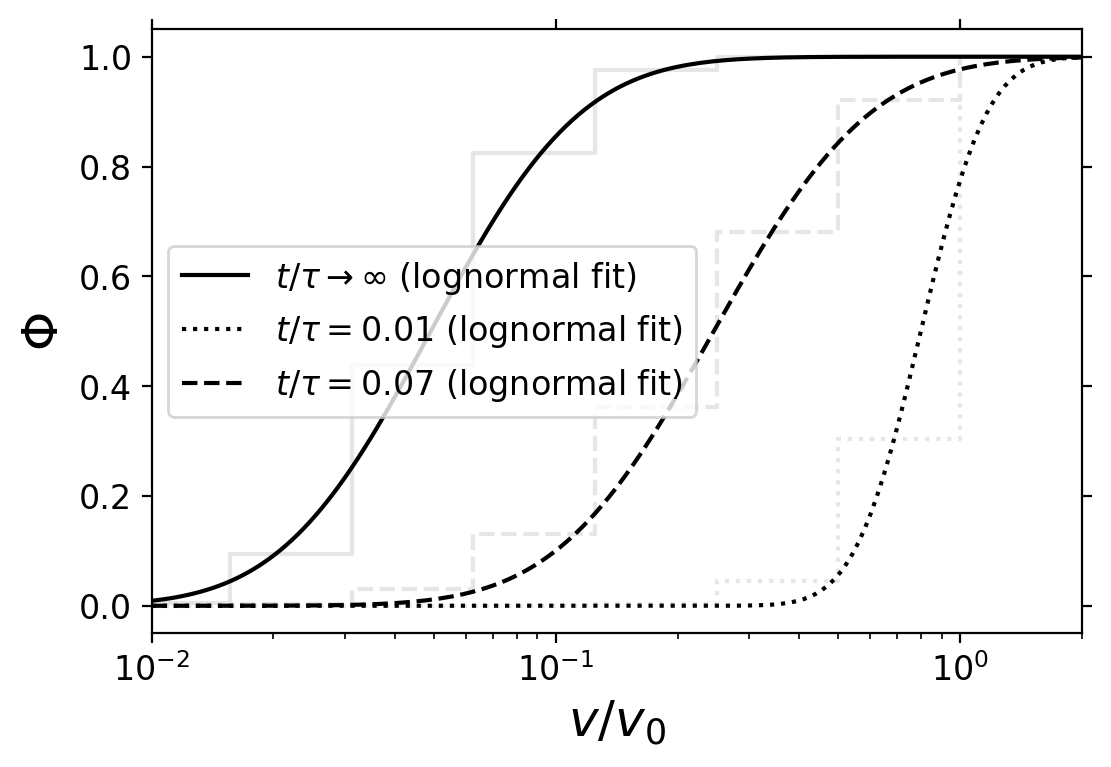}
\par\end{centering}
\caption{\label{fig:Phi_timeslice}For an initial strain of $\gamma=10$ and
a kinetic prefactor $\alpha=10$ (see figures \ref{fig:totalSA_t}
and \ref{fig:Phi_vary_alpha}), we show the time-evolution of the
cumulative distribution function (CDF) on the droplet size distribution,
$\Phi(v)=1/\phi\int_{0}^{v}n(v)vdv$. As droplets break apart, the
typical droplet size decreases and the variance in the distribution
increases.}
\end{figure}

In reality, there is no reason to expect such step-wise changes from
experimental observations: the delta-distributed size distribution
is a combined consequence of assuming a delta-distributed initial
condition and strict binary, symmetric breaking events. However, the
broad trends underlying the discrete distributions may in fact be
represented in experiments. In Figure \ref{fig:Phi_timeslice}, we
show that the broad trends reflected in the step-wise daughter droplet
distribution can be reasonably approximated by constructing a log-normal
distribution fit to the same mean and variance. At time-slices $t/\tau=0.01,0.07,\infty$,
we show the discrete CDF (light grey lines) and its smooth approximation
(black lines).

\begin{figure}
\begin{centering}
\includegraphics[scale=0.6]{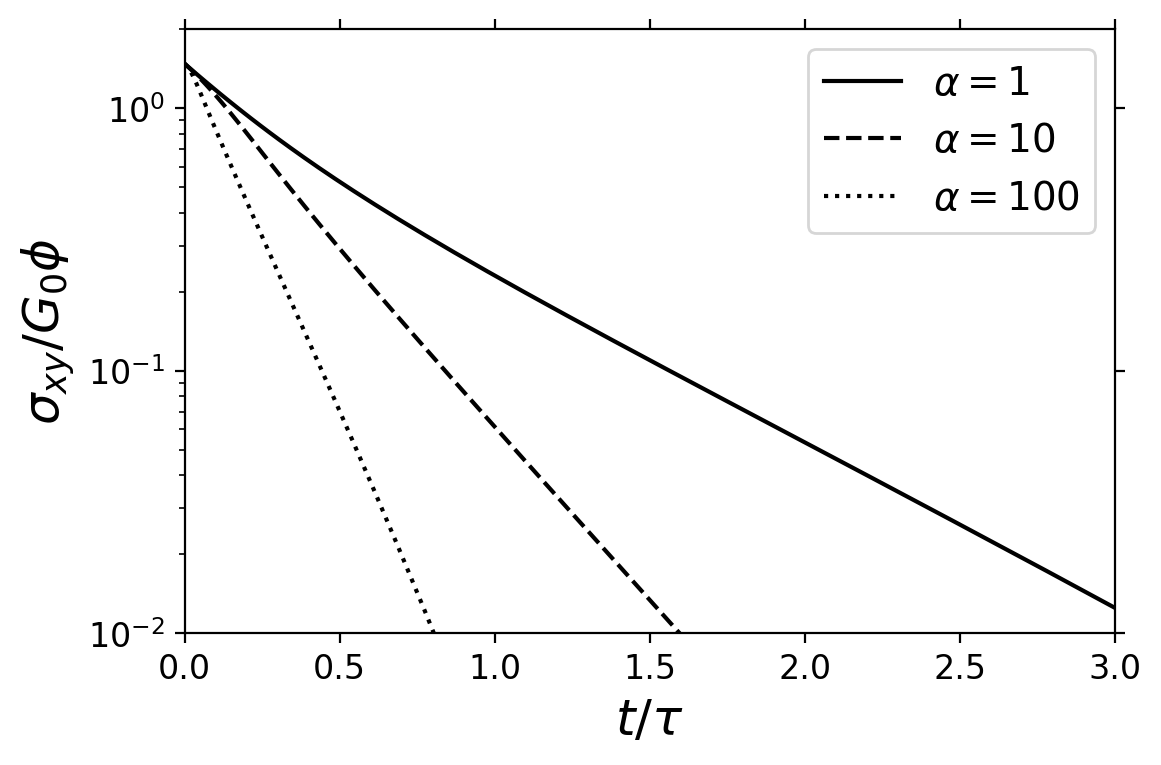}
\par\end{centering}
\caption{\label{fig:nlin_stress_relax}Predictions of the STEPB model for stress
relaxation following a strain of $\gamma=10$ with varying kinetic
prefactor to breakup $\alpha=1,10,100$ and $\zeta=0$. Accelerating
droplet breakup moves the droplet stress to faster-relaxing structures,
which in turn leads to faster stress relaxation dynamics. By comparing
stress relaxation kinetics at low strains (c.f. figure \ref{fig:linear_relaxn})
and high strains, it should be possible to infer an appropriate choice
of $\alpha$.}
\end{figure}

Figure \ref{fig:Phi_timeslice} is interesting to consider, but unfortunately
it may generally be difficult to directly measure a droplet size distribution
at intermediate relaxation times. Fortunately, however, signatures
of droplet breakup are also evident during stress relaxation. As droplets
break apart, the smaller daughter droplets are able to relax their
shape faster than the original parent droplets. In figure \ref{fig:nlin_stress_relax},
we confirm that the faster shape relaxation likewise translates to
faster stress relaxation, which is more readily measured than a droplet
size distribution.

\begin{figure}
\centering{}\includegraphics[scale=0.6]{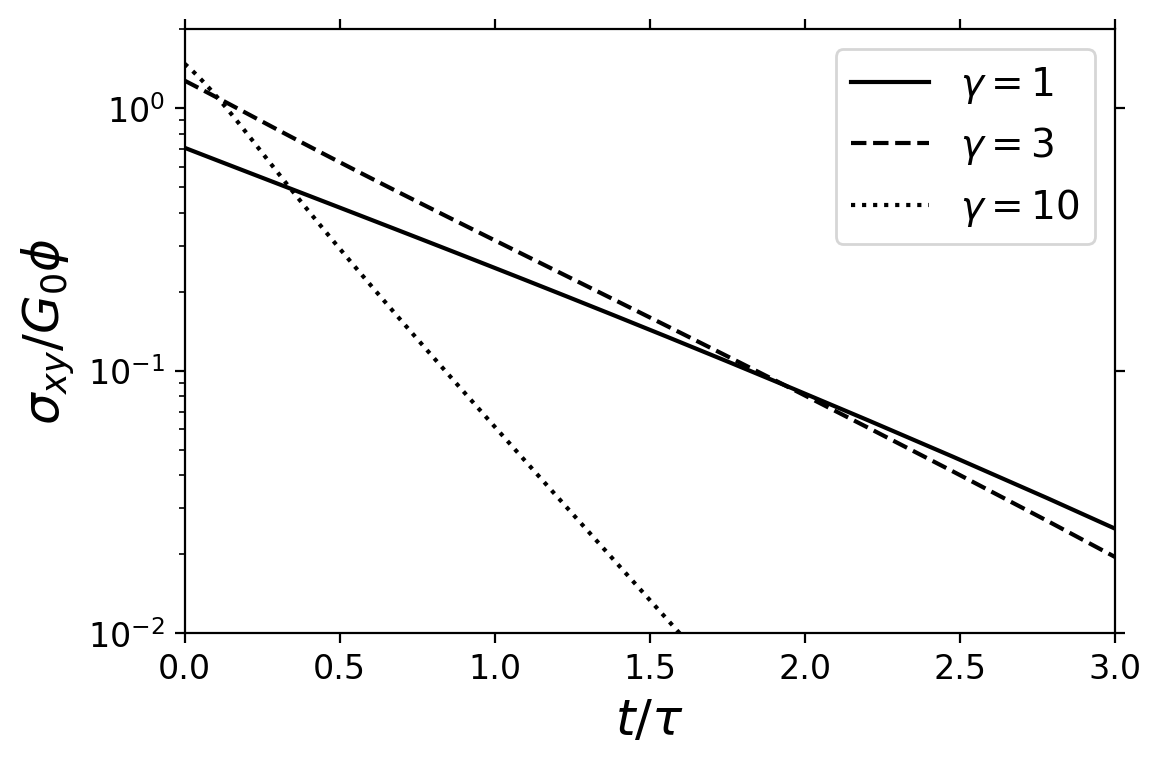}\caption{\label{fig:nlin_relax_strain_sweep}STEPB model predictions for stress
relaxation after strains of amplitude $\gamma=1,3,10$. The initial
stress increases with increasing strain, but after a short time the
droplets break apart and the resultant daughter droplets are able
to relax their stress more quickly, which translates to a lower overall
stress after a short time.}
\end{figure}

As a counterpart to figure \ref{fig:nlin_stress_relax}, increasing
strain amplitude can have a similar effect to increasing $\alpha$,
in that it produces faster breakup and smaller final droplet sizes.
Therefore, in figure \ref{fig:nlin_relax_strain_sweep} we show the
influence of strain amplitude on stress relaxation for fixed $\alpha=10$
at strains $\gamma=1,5,20$. For sufficiently small strains $\gamma=1$,
stress relaxation is effectively linear as previously seen in figure
\ref{fig:linear_relaxn}. For larger strains, stress relaxation quickly
becomes linear but with a characteristic relaxation time that decreases
with increasing strain amplitude, since the typical droplet size decreases
with increasing strain amplitude.

\subsection{\label{subsec:Strain-Sweep-to}Strain-dependence in final daughter-droplet
distribution}

\begin{figure}
\begin{centering}
\includegraphics[scale=0.6]{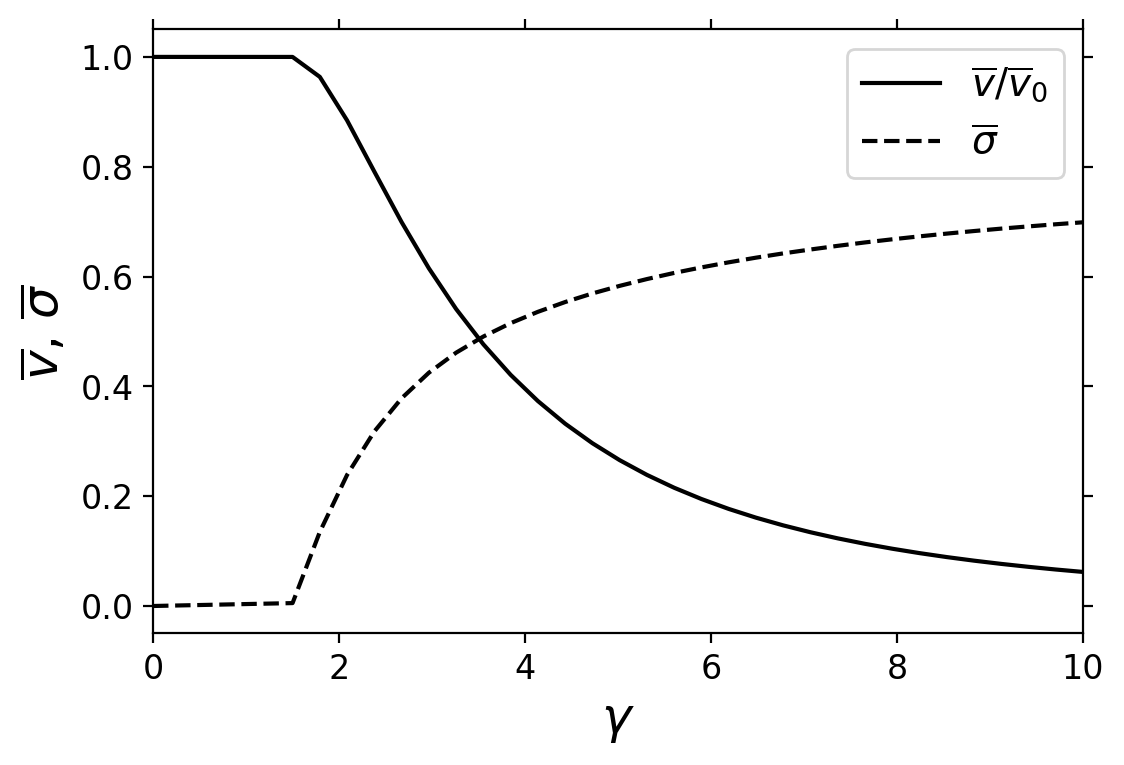}
\par\end{centering}
\caption{\label{fig:Strain_Sweep}Predictions of the STEPB model for the mean
and variance of the final daughter droplet distribution after an initial
step-strain deformation. Below a strain of about $\gamma=1.5$, droplets
are not predicted to break. Above that limit, however, increasing
initial strain leads to a smaller droplet size $\bar{v}/v_{0}$ and
higher relative variance $\bar{\sigma}$ in the distribution.}
\end{figure}

To close this section, we will discuss the broad trends that emerge
in STEPB when exploring the influence of initial strain on the final
daughter droplet distribution. Because changes in the size distribution
are reasonably approximated by a log-normal distribution (c.f. figure
\ref{fig:Phi_timeslice}), we will report these results in terms of
changes to the mean droplet size $\bar{v}/v_{0}$ and the relative
standard deviation $\bar{\sigma}=\langle v-\bar{v}\rangle^{1/2}/v_{0}$.
In figure \ref{fig:Strain_Sweep}, we see that for increasing initial
strain, the final daughter droplet distribution is characterized by
increasing polydispersity and decreasing average droplet size.

Although experimental data is not yet available for comparison, we
expect that the trends reflected in figure \ref{fig:Strain_Sweep}
are physically reasonable and we look forward to an eventual confrontation
with experimental data as and when such data becomes available. Here,
we once again remind the reader that our predictions for strain-dependent
daughter droplet distributions come from a very simple set of rules
governing the breakup process. There is ample room for refinement
as supported by data and required by application.

\begin{figure}
\begin{centering}
\includegraphics[scale=0.6]{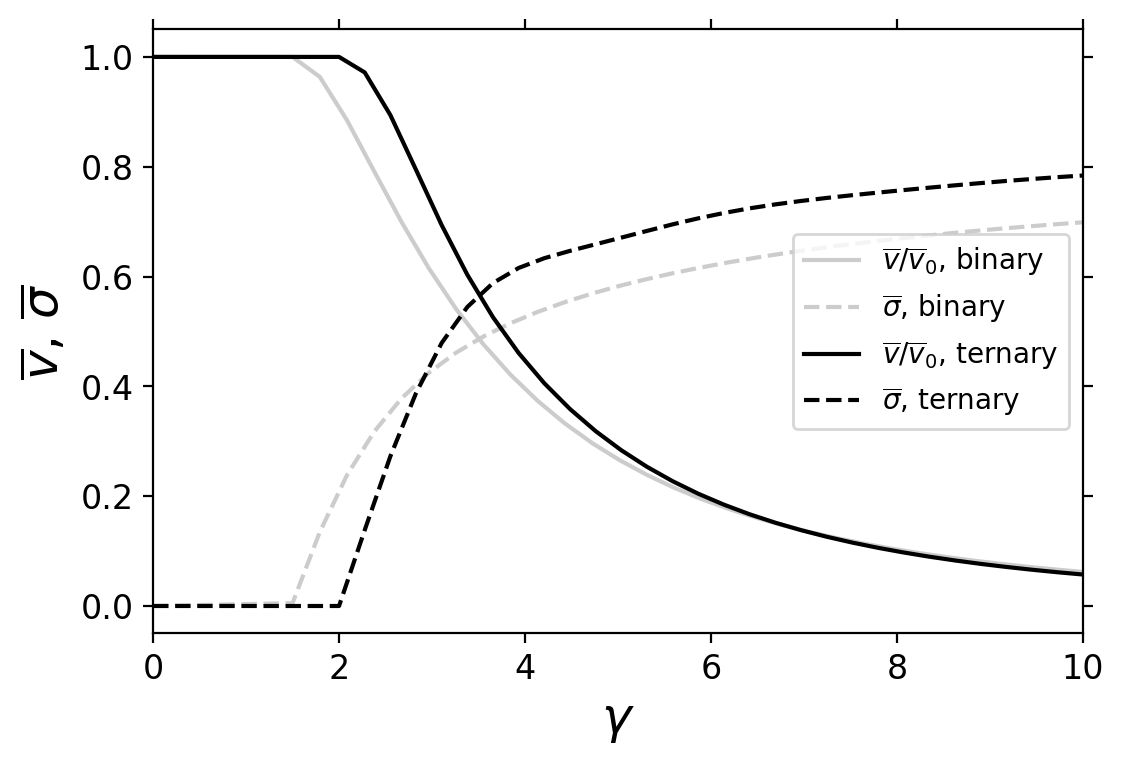}
\par\end{centering}
\caption{\label{fig:strain_sweep_ternary}Comparing predictions of the STEPB
model with binary/ternary breakup rules. Here, we consider the mean
and variance of the final daughter droplet distribution after an initial
step-strain deformation. Ternary breakage requires a higher initial
strain, but for sufficiently large strain the final droplet size is
roughly independent of the binary/ternary breakage rule. At high strains,
variance in the final droplet size distribution is greater for the
ternary breakup rule.}
\end{figure}

To close this section, we want to discuss the influence of the binary
breakage approximation, which is perhaps the most easily identifiable
weakness of the STEPB model. While it is possible to generalize the
STEPB model to consider arbitrary daughter droplet distributions,
here we will show that the predictions given in figure \ref{fig:Strain_Sweep}
are relatively insulated against changes to the breakup kernel that
would ordinarily be considered significant, even while all other model
parameters are held fixed. For example, the assumptions and approximations
of the STEPB model can easily be revised to ternary breakup, where
each droplet breaks apart into three identical daughter droplets with
the same deviatoric stress. As shown in figure \ref{fig:strain_sweep_ternary},
the move from binary to ternary breakup slightly increases the critical
strain, as the critical strain reflects a state where the broken droplets
(if spherical) have a lower surface area than the stretched droplet.
At higher strains, the final mean size of a daughter droplet $\bar{v}/v_{0}$
is roughly independent of the binary/ternary approximation, but the
variance in the distribution is slightly higher for ternary breakup.

Thus, while fine-tuning of the daughter droplet distribution will
have some effect on STEPB model predictions, such adjustments necessarily
introduce adjustable parameters at a cost to the model's simplicity.
As presently formulated for binary breakup, we feel that STEPB strikes
a good balance between physical detail and conceptual simplicity suitable
for many applications.

\section{\label{sec:experiments}Sample Calculations with Experimental Data}

The STEPB model was developed to address a gap in the existing emulsion
modeling literature, where droplet breakup in stabilized emulsions
is difficult to describe if the deformation is unsteady. In particular,
section \ref{sec:step-shear} looked at droplet breakup following
step-strain deformation, and it is our view that the STEPB model shows
tremendous promise in its ability to leverage simple approximations
for capturing complex phenomena relevant to droplet breakup in unsteady
deformations.

Outside of unsteady flow conditions, we will make no claims for the
conceptual or practical superiority of the STEPB model as presently
formulated. In steady simple shear flow, for example, the STEPB model
allocates a lot of resources to describing a droplet's shape and proclivity
to breakup - information that can be reasonably inferred from the
imposed shear rate alone. At the same time, STEPB allocates very few
resources to describing phenomena that are often relevant to the final
daughter droplet distribution in steady flow, namely (1) coalescence
(where applicable) and (2) complex daughter droplet distributions.

To reinforce this discussion, we will provide a comparison between
STEPB predictions and experimental results for oil/water emulsions
processed in a cone-mill device \cite{dubbelboer2016pilot}. The point
of this comparison is to highlight the weaknesses of the STEPB model
and to discuss why these weaknesses are less of a liability for the
unsteady deformations that STEPB is principally designed to address.

In a recent study by Dubbelboer et al \cite{dubbelboer2016pilot},
the authors prepared a series of dense emulsions (mayonnaise) from
oil, water, vinegar, egg yolk, and salt with varying oil concentration.
The components were initially mixed in a Silverson High Shear Mixer,
and the droplet size distribution was refined by high shear processing
in a cone mill. For an emulsion of 72\% oil, $\phi=0.72$, the authors
estimate a surface tension of $\Gamma=10$mN/m, an effective viscosity
of $\mu^{eff}=0.18$Pa$\cdot$s, and an initial average droplet diameter
of $37\mu$m. From these, we can estimate a typical droplet relaxation
time of $\tau=\mu^{eff}r_{0}/\Gamma=3.3\cdot10^{-4}s$. The cone mill
effectively exposes droplets to steady shear flow with a shear rate
of $\dot{\gamma}=25000/s$, which gives us an estimated overall Capillary
number of $Ca=8.5$ and we expect to see significant changes in the
droplet size distribution. Our STEPB predictions will assume $\zeta=0$,
and the initial/final droplet size distributions are lifted from figures
3b and 10a of the Dubbelboer report \cite{dubbelboer2016pilot}.

Numerical solutions of the STEPB model are obtained via the miCDF
scheme \cite{peterson2022miCDF}, the principle of which is covered
in more detail in appendix \ref{sec:Addition-Details-on} of the present
work. In summary, miCDF divides the droplet size distribution into
fractions of equal volume and tracks the typical droplet size/shape
in each fraction over time. In our view, miCDF is an extremely useful
discretization strategy for coupled rheology/PBE problems because
(1) large changes in the size distribution do not require re-meshing
the size distribution and (2) compared to other ``adaptive'' methods,
miCDF does not lose accuracy when handling a moving reference droplet
volume in the shape evolution equation.

\begin{figure}
\begin{centering}
\includegraphics[scale=0.6]{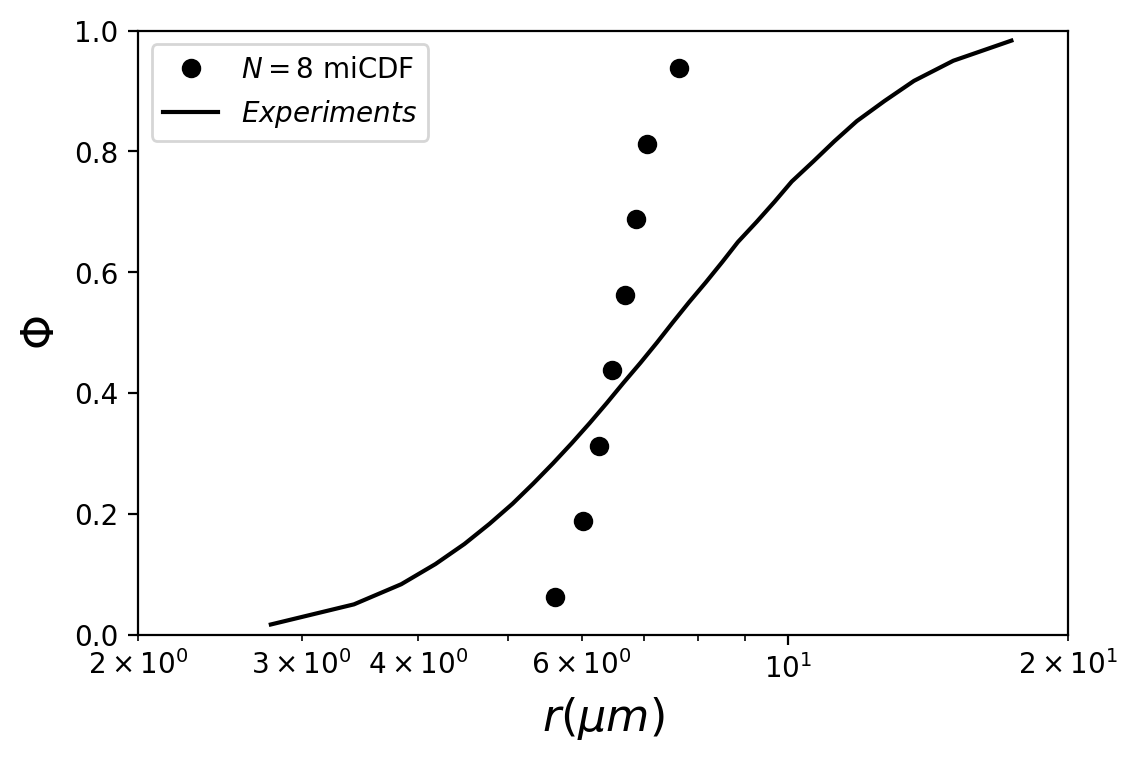}
\par\end{centering}
\caption{\label{fig:experiments}Comparing predictions from the STEPB model
against experimental observations for the steady state droplet size
distribution (in terms of a cumulative size distribution) for a stabilized
oil/water emulsion $\phi=0.72$ as described in work done by Dubbelboer
et. al. {[}REF{]}. Note that STEPB captures the typical droplet size
reasonably well, but dramatically underestimates the variance in the
final distribution.}
\end{figure}

Having established the context of our simulations, we now compare
STEPB predictions for the final droplet size distribution against
the results from Dubbelboer et. al. In figure \ref{fig:experiments},
we show that STEPB seems to capture the average droplet size reasonably
well, but dramatically underestimates the variance of the distribution.

The broader droplet size distribution observed in experiments shows
that the STEPB model is missing important physics relevant to steady
simple shear flows, but the general agreement in average droplet size
indicates that there are key ideas within STEPB worth building on.
In terms of weaknesses, the broader droplet size distribution could
be achieved by some combination of (1) directly assigning a broader
daughter droplet distribution and/or (2) allowing droplet coalescence
to occur. The latter of these two modes is particularly interesting:
by allowing coalescence, one can sustain a population of droplets
with size above the critical size for breakup. These larger droplets
can become highly stretched, facilitating successive generations of
droplet breakup and extending the droplet size distribution well below
the critical size for breakup at the imposed shear rate. However,
for stabilized dense emulsions with an excess of surfactant (such
as the mayonnaise used in these experiments) the mechanism for coalesence
is unclear, and so complex daughter droplet distributions are a more
likely explanation.

To summarize the above, we suggest that figure \ref{fig:experiments}
shows that STEPB accurately identifies a critical capillary number
as reflected by the typical droplet size surviving at steady state,
but fails to capture the breadth of the droplet size distribution
possibly due to the omission of coalescence in the present implementation.
However, implementing coalescence in the STEPB model is not obvious:
how does one interpret the process of coalescence in terms of ellipsoids
alone? If coalescence happens under deformation but not at equilibrium,
how should the shape tensor enter into the coalescence rate expression?
This is a target for future work, and may require higher-order tensor
representations of droplet shape evolution.

Having acknowledged these weaknesses, we remind the reader that the
current formulation of the STEPB model was designed to address flows
at high Deborah numbers (i.e. closer to step deformation than steady
shear). High Deborah number flows could be encountered in contractions/expansion
zones for processing equipment and valves, but also in fully resolved
turbulent flows for conventional mixers. At high Deborah numbers,
droplet breakup is dominated by the strain introduced in a narrow
interval of time, and a broad distribution of daughter droplets can
be found without direct fine-tuning of the daughter droplet distribution.
By contrast, at low Deborah numbers all droplets above a critical
size will eventually break, and virtually no droplets below that critical
size will break, such that the final daughter droplet distribution
bears a direct reflection of what one has assumed about the daughter
droplet distribution for individual breakup events.

Since there are already existing models to cover low Deborah number
applications \cite{dubbelboer2016pilot, maindarkar2014prediction},
poor performance of the STEPB model under steady shear does not seriously
detract from its overall usefulness, in our view. That being said,
it should be possible to improve the model so that it performs suitably
in both high and low Deborah number flows.

\section{\label{sec:Summary-and-Conclusions}Summary and Conclusions}

In this paper, we introduced the ``shape tensor emulsion population
balance'' (STEPB) model for coupling the evolution equations on droplet
shape and droplet size distribution. The main weaknesses of the model
as presently formulated are (1) it is restricted to interpolate the
breakup process as a transition from an ellipsoidal parent droplet
to a collection of ellipsoidal daughter droplets, (2) it neglects
terms related to droplet coalescence, and (3) it relies on a very
simple symmetric binary breakup approximation of the daughter droplet
size distribution. In spite of these weaknesses, we have shown that
the STEPB model yields non-trivial predictions regarding a complex
strain-dependent daughter droplet distribution under step-strain deformations.
This is a result that would challenge competing emulsion models, and
in STEPB it is accomplished with only four parameters; a surface tension
$\Gamma$, a relaxation time $\tau$, a ``slip'' parameter $\zeta$,
and a kinetic prefactor to breakup $\alpha$. The slip parameter $\zeta$
accounts for non-affine deformation as might occur if the dispersed
droplet were of a much higher viscosity that the surrounding medium.
Like the viscosity ratio, the slip parameter also controls the critical
capillary number in simple shear flow (c.f. appendix \ref{sec:Slip-and-the}).
The kinetic prefactor $\alpha\sim O(1)$ ties the processes of breakup
and shape relaxation together, as needed to ensure that the joint
relaxation/breakup processes always decrease total surface area.

Looking towards future research with the STEPB modeling framework,
there are three main directions that seem worth pursuing in terms
of improvements upon the STEPB model itself. In a first direction,
there are opportunities to incorporate additional physics into the
STEPB model: we can move towards higher-order tensors, readmit droplet
coalescence, fine-tune the daughter droplet distribution, add memory
into the droplet breakup rate, and so on. In a second direction, by
contrast, there are opportunities to pursue further simplifications:
we could perhaps use the STEPB model as a benchmark for evaluating
the performance of reduced-order models, with complexity in-line with
a generalized Doi-Ohta model. Finally, a third research direction
(possibly in conjunction with the second) would be to embed the STEPB
model (or a simplified variant) into CFD model calculations to better
understand and optimize the processes by which dense emulsions are
formed in industry applications.

Besides these modeling directions, the STEPB model opens up new opportunities
for experimental and computational probing of droplet breakup dynamics
in dense emulsions. Most notably, droplet breakup in step deformation
appears to be a useful and well-defined problem to consider. It would
be interesting to develop experiments or simulations that allow for
simultaneous real-time measurement of droplet deformation, stress
relaxation, and breakup of dense suspensions at high Capillary numbers.

In all, it is our view that the STEPB model represents a major conceptual
advance in the available tools for modeling droplet breakup in dense
emulsions and there are many promising directions for follow-up work,
through both experimental and computational/theoretical endeavors.

\section{Acknowledgments}

The authors would like to acknowledge Unilever and the CAFE4DM consortium
for funding and technical input. We also acknowledge financial support
from the European Research Council under the Horizon 2020 Programme,
ERC grant agreement number 740269.

\section*{References}
\bibliography{references}
\appendix

\section{\label{sec:Addition-Details-on}Addition Details on Numerical Methods}

In section \ref{sec:step-shear}, we will consider calculations for
droplets that are initially monodisperse, $n(t=0,v)=\delta(v-v_{0})$,
in which case equation \ref{eq:ndot_final} yields a solution of the
form:

\begin{equation}
n(t,v)=\sum_{i=0}^{\infty}n_{i}(t)\delta(v-v_{0}/2^{i})\label{eq:delta_sum}
\end{equation}

where the droplets are delta-distributed at volumes reflecting successive
divisions of the starting droplet. Inserting equation \ref{eq:delta_sum}
into equation \ref{eq:ndot_final} and grouping terms at each delta-peak
yields the following set of evolution equations:

\begin{equation}
\ppt n_{0}=-g_{0}n_{0}
\end{equation}

\begin{equation}
\ppt n_{i>0}=-g_{i}n_{i}+2g_{i-1}n_{i-1}
\end{equation}

where $g_{i}=g(v_{i})$. For our purposes, we truncate equation \ref{eq:delta_sum}
when droplets are small enough that no further breaking takes place
within the model.

If the initial droplet size distribution is not delta-distributed,
then choosing an appropriate discretization strategy is more difficult.
In principle, one could break the initial condition into a finely-resolved
set of delta-peaks and repeat the process outlined above, but this
will typically be inefficient when the eventual goal is to embed the
emulsion model into CFD calculations.

While many discretization strategies may be possible, we will employ
our recently devised ``method of the Inverse Cumulative Distribution''
(miCDF) scheme, which effectively breaks the population into pre-allocated
intervals of total volume, and then tracks the median droplet size
within each interval. To expand on this summary, we first define a
volume density distribution $\phi(v)=vn(v)$ as a simple volume-weighted
transformation of the number density distribution. This volume density
distribution is conservative, $\int_{0}^{\infty}dv\phi(v)=\phi_{0}$
and as such can be interpreted as a probability density function.
From that probability density function, we can define a cumulative
distribution function (CDF) $\Phi(v)$:

\begin{equation}
\Phi(v)=\frac{1}{\phi_{0}}\int_{0}^{v}dv'\phi(v)
\end{equation}

For a given droplet size $v$, the CDF $\Phi$ gives the total volume
fraction of droplets with size less than $v$. Conversely, the inverse
CDF, $v(\Phi)$ gives the limiting droplet size needed to contain
a specified fraction $\Phi$ of the total droplet volume fraction.
Note that because $\phi(v)>0$, $\Phi(v)$ is monotonically increasing
in $v$, as needed for the inverse function to exist. An evolution
equation for $v(\Phi)$ is produced by applying the triple product
rule:

\begin{equation}
\Big(\frac{\partial v}{\partial t}\Big)_{\Phi}=-\Big(\frac{\partial v}{\partial\Phi}\Big)_{t}\Big(\frac{\partial\Phi}{\partial t}\Big)_{v}=-\frac{1}{\phi}\int_{0}^{v}dv'\ppt\phi(v)
\end{equation}

Recasting the evolution equation in terms of the inverse CDF can be
advantageous whenever there are large changes in the droplet size
distribution, but it is not the only method with this property \cite{McGraw1997a, Vikhansky2015}.
Where the PBE is coupled to an underlying viscoelastic constitutive
equation, however, it is our view that the miCDF discretization strategy
is a more natural and more flexible strategy overall.

When the droplet shape is tracked along moving reference droplet size
$v^{*}(t)$, the shape evolution equations \ref{eq:shape_PBE3} and
\ref{eq:shape_PBE4} must contain an additional ``advection'' term
to account for changes in shape that occur simply due to changes in
the reference droplet size:

\begin{equation}
\Big(\frac{\partial\Tb}{\partial t}\Big)_{v^{*}}=\cdots+\Big(\frac{\partial\Tb}{\partial v}\Big)_{v=v^{*}}\Big(\frac{\partial v^{*}}{\partial t}\Big)\label{eq:moving_ref_Theta}
\end{equation}

where the ellipses denote terms previously derived for the fixed droplet
volume reference frame (deformation, relaxation, population balances,
etc). A moving reference frame of this nature is common to most of
the PBE methods considered in coupled CFD/PBE applications, including
those based on the popular ``Quadrature Method of Moments'' QMOM
approach \cite{McGraw1997a, Vikhansky2015}. In QMOM, moment evolution
equations are closed via an expensive moment inversion approximation
that can be accurate for quadrature in $v$ but not especially useful
for evaluating derivatives in $v$, as needed for the moving reference
frame in equation \ref{eq:moving_ref_Theta}. By contrast, miCDF avoids
the slow moment inversion step and yields a consistent order of accuracy
across both number density and shape evolution equations. In our calculations,
we rely on simple linear interpolations and trapezoid quadrature schemes
for numerical calculations.

\section{\label{sec:Slip-and-the}Slip and the Critical Capillary Number}

\begin{figure}
\begin{centering}
\includegraphics[scale=0.6]{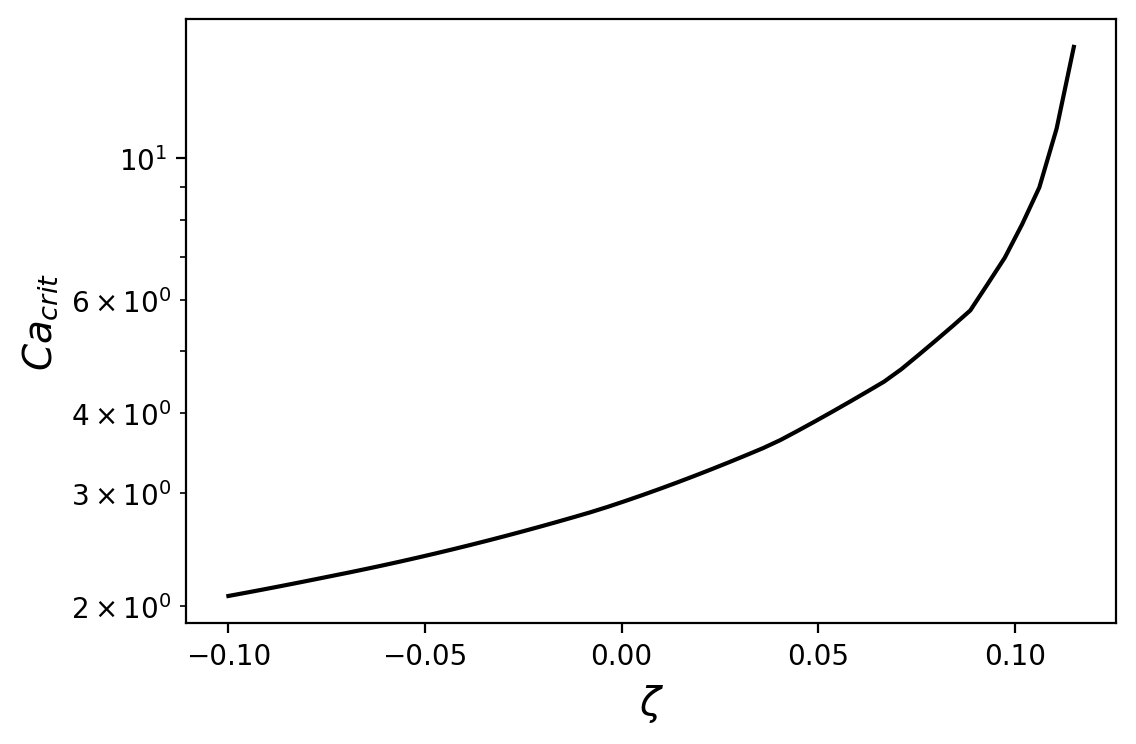}
\par\end{centering}
\caption{\label{fig:ca_crit}Predictions for the critical capillary number
$Ca_{crit}$ of the STEPB model as a function of the slip parameter
$\zeta$. With increasing slip, higher shear rates are needed to induce
breakup in simple shear; for slip exceeding $\zeta>0.11$, droplet
breakup cannot be achieved at any shear rate.}
\end{figure}

In the STEPB model, we describe droplets as ellipsoids that deform
non-affinely in flow with a uniform ``slip'' factor $\zeta$ subtracted
against the bulk strain gradient. This slip is partially intended
to reflect the influence of varying relative viscosity of the droplet
itself and the surrounding medium - where the surrounding medium is
less viscous, for example, droplets will resist deformation relative
to the background flow {[}REF{]}. In dilute emulsions, for example,
the ``critical capillary number'' $Ca_{crit}$ for droplet breakup
depends on this viscosity ratio, and even diverges for a viscosity
ratio of $4$. In keeping with that, we show in figure \ref{fig:ca_crit}
that $Ca_{crit}$ diverges for $\zeta\sim0.11$ in the STEPB model.
This analogy breaks down for negative values of $\zeta$, where one
might hope to represent a low viscosity dispersed phase: in reality,
$Ca_{crit}$ is a non-monotonic function of the viscosity ratio, with
intermediate viscosity ratios having the lowest $Ca_{crit}$. This
is a potential liability for the STEPB model, but for dense suspensions
the effective viscosity in the medium surrounding a droplet is not
necessarily the viscosity of the continuous phase.

\end{document}